\documentclass[
reprint,
showkeys,
longbibliography,
nofootinbib,
superscriptaddress,
aps,
notitlepage
]{revtex4-1}
\usepackage[utf8]{inputenc}
\usepackage{mathtools}
\usepackage{amsmath}
\usepackage{amssymb}
\usepackage[colorlinks,unicode]{hyperref}
\usepackage[capitalise]{cleveref}
\usepackage[dvipsnames]{xcolor}
\usepackage{graphicx}
\usepackage{multirow}
\usepackage{bm}
\usepackage{physics}
\usepackage{siunitx}
\usepackage{fontawesome} 
\usepackage{placeins}

\usepackage{lineno}

\usepackage[caption=false]{subfig}

\usepackage{tikz}
\usepackage[compat=1.1.0]{tikz-feynman} 

\makeatletter
\renewcommand\onecolumngrid{%
\do@columngrid{one}{\@ne}%
\def\set@footnotewidth{\onecolumngrid}%
\def\footnoterule{\kern-6pt\hrule width 1.5in\kern6pt}%
}
\makeatother

\newcommand\myshade{80}
\colorlet{mylinkcolor}{ForestGreen}
\colorlet{mycitecolor}{Red}
\colorlet{myurlcolor}{violet}
\usepackage[dvipsnames]{xcolor}

\hypersetup{
  linkcolor  = mylinkcolor!\myshade!black,
  citecolor  = mycitecolor!\myshade!black,
  urlcolor   = myurlcolor!\myshade!black,
  colorlinks = true
}

\usepackage{enumitem}
\newlist{todolist}{itemize}{2}
\setlist[todolist]{label=$\square$}
\usepackage{pifont}
\DeclareSIUnit\solarmass{\ensuremath{\mathrm{M}_\odot}}
\DeclareSIUnit\parsec{pc}
\DeclareSIUnit\year{yr}

\newcommand{\DP}{\gamma^\prime}
\newcommand{\mDP}{m_{\DP}}
\newcommand{\diff}{\mathrm{d}}

\newcommand{\IFCA}{Instituto de F\'isica de Cantabria (IFCA, UC-CSIC), Avenida de
Los Castros s/n, 39005 Santander, Spain}

\begin{document}

\preprint{}

\title{Cosmology and direct detection of the Dark Axion Portal}

\author{Juan Cortabitarte Guti\'errez}
\email{juan.cortabitarteg@alumnos.unican.es}
\affiliation{\IFCA}

\author{Bradley J. Kavanagh}
\email{kavanagh@ifca.unican.es}
\affiliation{\IFCA}

\author{N\'uria Castell\'o-Mor}
\email{nuria.castello.mor@cern.ch}
\affiliation{\IFCA}

\author{Francisco J. Casas}
\email{casas@ifca.unican.es}
\affiliation{\IFCA}

\author{Jose M. Diego}
\email{jdiego@ifca.unican.es}
\affiliation{\IFCA}

\author{Enrique Martínez-González}
\email{martinez@ifca.unican.es}
\affiliation{\IFCA}

\author{Roc\'{\i}o Vilar Cortabitarte}
\email{vilar@ifca.unican.es}
\affiliation{\IFCA}


\begin{abstract}
The Dark Axion Portal provides a model for Dark Matter (DM) in which both Dark Photons $\DP$ and Axions $a$ can contribute to the present day abundance of DM. We study the parameter space of the Dark Axion Portal to pinpoint regions of the parameter space where $\DP$ and $a$ can be produced with sufficient abundance to account for the cosmic DM density, while still being detectable in planned direct detection and axion haloscope experiments. In particular, we explore the production of eV-scale Dark Photons in the Dark Axion Portal, taking into account a possible kinetic mixing between the dark and visible photons, which is essential for the detection of dark photons through absorption in direct searches. We show that a non-zero kinetic mixing does not generally spoil the phenomenology of the model, leaving both the axion and dark photon stable. Viable production mechanisms point to a sub-dominant population of dark photons making up $\lesssim 10\%$ of the DM, with the remainder consisting of axion DM. Dark photons in the mass range $\mDP \sim 20-200\,\mathrm{eV}$ and axions in the mass range $m_a \sim 30 - 400\,\mu\mathrm{eV}$ may be produced with these abundances self-consistently in the Dark Axion Portal and are within the reach of future direct searches.
\href{https://github.com/bradkav/DarkAxionPortal}{\faGithub} \href{https://doi.org/10.5281/zenodo.5794543}{\faTags}
\end{abstract}

\maketitle

\section{Introduction}

The nature and properties of Dark Matter (DM) are among the greatest outstanding problems in astrophysics, particle physics and cosmology~\cite{Bertone:2004pz,deSwart:2017heh}. An extensive campaign is underway to try to detect DM, both in laboratory experiments and using astrophysical observations. Weakly Interacting Massive Particles (or WIMPs)~\cite{Roszkowski:2017nbc} are one of the leading candidates to explain the DM particle. However, there are as many search strategies as there are well-motivated candidates, which include sterile neutrinos~\cite{Boyarsky:2018tvu}; pseudo-scalar particles known as axions~\cite{Marsh:2015xka}; gauge bosons of a new `Dark' $U(1)$ symmetry, referred to as Dark Photons or Hidden Photons~\cite{Caputo:2021eaa}; and even massive compact objects such as primordial black holes~\cite{Green:2020jor}. In particular, the failure to detect WIMPs at the GeV--TeV scale~\cite{Blanco:2019hah} has lead to an increasing interest in such a broader range of candidates~\cite{Bertone:2018krk}.

One possibility is that there may be a more complicated `Dark Sector', comprised of many new particles and forces~\cite{Pospelov:2007mp,Essig:2013lka,Alexander:2016aln}. To evade detection so far, this Dark Sector must be only weakly coupled to the visible sector. Attempts have been made to generalise the ways in which these two sectors may interact, classifying these into different \textit{portals}, depending on which particles and interactions link the two sectors. These include the Higgs portal~\cite{Arcadi:2019lka}, the vector portal~\cite{Fortuna:2020wwx}, the sterile neutrino portal~\cite{Escudero:2016tzx,Escudero:2016ksa} and the axion portal~\cite{Nomura:2008ru}. Not only do these portals offer a more systematic way of studying DM, they also give rise to novel experimental signatures.

In this work, we study the Dark Axion Portal, introduced in Ref.~\cite{Kaneta:2016wvf}. The Dark Sector of the Dark Axion Portal includes two particles which may contribute to the density of DM in the Universe: a pseudo-scalar QCD axion $a$ and a vector dark photon $\DP$. The QCD axion provides a compelling solution to the Strong CP problem~\cite{DiLuzio:2020wdo}, while the dark photon is a simple extension of the Standard Model which has been invoked to explain a number of anomalies, including measurements of the muon g-2~\cite{Kirpichnikov:2020tcf,Filippi:2020kii,Ge:2021cjz}. The presence of both particles gives rise to new interactions which enrich the phenomenology of the model, not least because in this case the DM may be comprised of not one but two particle species~\cite{Zurek:2008qg}.

The Dark Axion Portal has been studied in a number of contexts previously, including possible astrophysical constraints~\cite{Choi:2018dqr,Choi:2018mvk,Choi:2019jwx,Ratzinger:2020oct,Arias:2020tzl,Hook:2021ous,Nakayama:2021avl} and proposed collider searches~\cite{Alvarez:2017eoe,deNiverville:2018hrc,deNiverville:2019xsx,Biswas:2019lcp,Deniverville:2020rbv}, with a focus on keV-scale dark photons and heavier. The connection between the axion and the dark photon may also lead to novel production mechanisms for the two particles. For example, production of dark photons may be dominated by freeze-in processes~\cite{Hall:2009bx} which are mediated by the axion. These scenarios, where the dark photon has no kinetic mixing with the photon, were studied in detail in Ref.~\cite{Kaneta:2016wvf,Kaneta:2017wfh}, and can give rise to the correct relic DM density in parts of the parameter space of the model. The axion density may also be affected by these couplings. Indeed, the decay of the axion field into the dark gauge field may lead to a dilution of the axion energy density, preventing overclosure in scenarios with, for example, very large values of the axion decay constant $f_a$~\cite{Agrawal:2017eqm,Kitajima:2017peg,Agrawal:2018vin,Co:2018lka,Caravano:2021bfn,Kakizaki:2021mgj}. 
The Dark Axion Portal may also provide a mechanism for cosmological relaxation~\cite{Domcke:2021yuz}, allowing the model to solve both the DM problem and also potentially the hierarchy problem of the Weak Scale.

From a cosmological perspective, multi-component DM scenarios are very promising in solving long standing problems present in the standard $\Lambda$-Cold Dark Matter ($\Lambda$CDM) cosmology at galactic and sub-galactic scales. These small-scale problems include the missing satellite problem, the too-big-to-fail problem and the core-cusp problem~\cite{DelPopolo:2016emo}, which could perhaps be alleviated by a mixture of cold DM and warm or hot DM which suppresses structure on small scales~\cite{2012JCAP...10..047A,Todoroki_2018,Parimbelli:2021mtp}. As we will see, for certain masses and couplings, the Dark Axion Portal provides such a mixed DM scenario.
Sufficiently light axions and axion-like particles may also contribute to solving these small-scale problems as they naturally predict a soliton in the central regions of galaxies; a minimum threshold mass for haloes~\cite{Schive2014}; and in addition can explain flux anomalies observed in lensed quasars~\cite{Amruth2022}.

In this work, we explore the cosmology and detectability of the Dark Axion Portal, with specific focus on scenarios in which both the dark photon and axion may contribute appreciably to the local DM density and in which both may be detectable in current and future DM searches. Such a dual detection would provide a crucial confirmation of the underlying model of DM and highlights the complementarity of different search strategies. For the dark photon, we consider the possibility of detection through absorption in direct detection experiments, giving rise to an ionisation signal~\cite{An:2014twa,Hochberg:2016sqx,Bloch:2016sjj}. The ionisation energy in typical target materials points us towards dark photons with masses in the range $\mDP \sim 1 - 1000\,\mathrm{eV}$. Crucially, this requires us to include a kinetic mixing between the dark and visible photons, which has previously received little attention in studies of the Dark Axion Portal. For the axion, we consider the sensitivity of terrestrial haloscopes, which search for axion-to-photon conversion for axion masses in the range $m_a \in [10^{-6}, 10^{-3}]\,\mathrm{eV}$, with the precise mass depending on the experimental set-up~\cite{Sikivie:1983ip,Semertzidis:2021rxs}.

In Sec.~\ref{sec:DarkAxionPortal}, we will summarise the Dark Axion Portal model and the specific Dark KSVZ realisation which we focus on. We then go on to discuss the cosmological production of both the axion and dark photon in Sec.~\ref{sec:Production}, including possible contributions from the kinetic mixing of the dark photon. In Sec.~\ref{sec:Detection}, we discuss the prospects for detecting the two DM candidates in axion haloscopes and low-threshold direct detection experiments. Finally, in Sec.~\ref{sec:Discussion}, we discuss the implications of these results and highlight the regions of parameter space where a consistent model including both particles could be conclusively detected in upcoming experiments. A summary of these results is given in Figs.~\ref{fig:Complementarity_10pct}~and~\ref{fig:Complementarity_1pct}.

Code for reproducing the results and plots of this paper is publicly available online at \href{https://github.com/bradkav/DarkAxionPortal}{github.com/bradkav/DarkAxionPortal}~\cite{DarkAxionPortalCode}.


\section{Dark Axion Portal}
\label{sec:DarkAxionPortal}

\begin{figure}[tb]
    \centering
    \includegraphics[trim={0.2cm 1.3cm 0.2cm 1.4cm}, clip, width=0.45\textwidth]{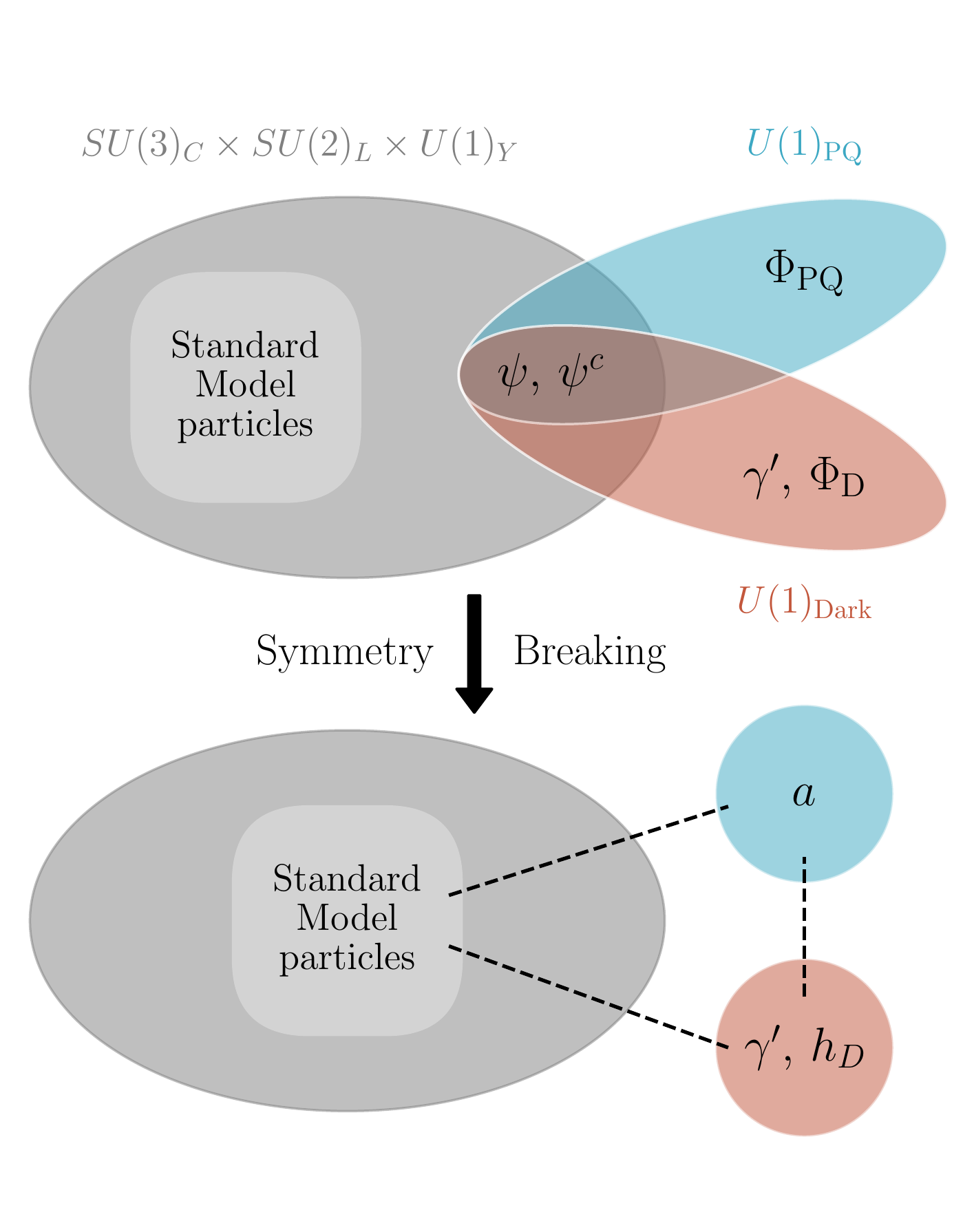}
    \caption{\textbf{Overview of the Dark Axion Portal Model}. We introduce a dark photon $\gamma^\prime$ and a Dark complex scalar $\Phi_D$, charged under a Dark $U(1)_D$ gauge symmetry, and a Peccei-Quinn field $\Phi_\mathrm{PQ}$, charged under a global $U(1)_\mathrm{PQ}$. New heavy fermions $\psi$ and $\psi_c$ are charged under these new symmetries, as well as (a subset of) the Standard Model symmetries. After the breaking of the $U(1)_D$ and $U(1)_\mathrm{PQ}$, we are left with a massive dark photon $\gamma^\prime$ and Dark Higgs $h_D$, as well as a light axion $a$. Interactions between the different sectors are mediated by loops of heavy fermions, as well as kinetic mixing $\epsilon$ between the Dark and visible photons.}
    \label{fig:DarkAxionPortal_illustration}
\end{figure}

In the Dark Axion Portal, a global Peccei-Quinn symmetry $U(1)_\mathrm{PQ}$ and a dark gauge symmetry $U(1)_D$ are added to the Standard Model (SM), along with a number of new fields. Here, we consider a version of the Dark KSVZ model proposed in Ref.~\cite{Kaneta:2016wvf}, inspired by the original  Kim-Shifman-Vainshtein-Zakharov axion model~\cite{Kim:1979if,Shifman:1979if}\footnote{Further details, and a broad review of possible QCD axion models, can be found in Ref.~\cite{DiLuzio:2020wdo}.}. The new fields include a dark photon $\DP$, two Higgs singlets $\Phi_\mathrm{PQ}$ and $\Phi_D$, and two heavy Weyl fermions $\psi$ and $\psi^c$.  A schematic illustration of the model is shown in Fig.~\ref{fig:DarkAxionPortal_illustration}

The $U(1)_\mathrm{PQ}$ symmetry is spontaneously broken at an energy scale $f_a$, known as the axion decay constant. The heavy fermions $\psi$ and $\psi^c$ couple to $\Phi_\mathrm{PQ}$ through a Yukawa interaction, such that after the $U(1)_\mathrm{PQ}$ symmetry breaking, $\psi$ and $\psi^c$ obtain a large mass $m_\psi \sim f_a$. The axion $a$ is identified as the CP-odd angular component of the field $\Phi_\mathrm{PQ}$ which, after the QCD phase transition, gains a mass~\cite{DiLuzio:2020wdo}:
\begin{equation}
    m_a \approx \frac{\sqrt{z}}{1+z} \frac{f_\pi}{f_a}m_\pi \approx 6\,\mu\mathrm{eV}\left(\frac{10^{12}\,\mathrm{GeV}}{f_a}\right)\,,
    \label{eq:axionmass}
\end{equation}
where $z = m_u/m_d \approx 0.56$, $m_\pi \approx 135\,\mathrm{MeV}$ is the pion mass and $f_\pi \approx 92 \,\mathrm{MeV}$ is the pion decay constant. 

The Dark scalar field $\Phi_D$ is given a potential of the form:
\begin{equation}
    V(\Phi_D) = -\mu^2 |\Phi_D|^2 + \lambda  |\Phi_D|^4 + \ldots\,,
\end{equation}
which leads to the development of a vacuum expectation value for $\Phi_D$, $\langle \Phi_D \rangle = \mu/\sqrt{2\lambda} \equiv v_D/\sqrt{2}$. Through this spontaneous breaking of the $U(1)_D$ symmetry, the dark photon gains a mass  $\mDP^2 = e'^2 D_{\Phi_D}^2 v_D^2$, where $e'$ is the Dark $U(1)_D$ coupling constant, and the radial mode of $\Phi_D$ becomes a Dark Higgs, with mass $m_{h_D} = \sqrt{2\lambda}v_D$~\cite{Mondino:2020lsc}. We will assume that $m_{h_D} > \mDP$.

The particles of the Standard Model are uncharged under the new $U(1)_\mathrm{PQ}$ and $U(1)_D$ symmetries. However, communication between the dark and visible sectors is achieved in a number of ways. The dark photon will mix with the Standard Model photon\footnote{We will be interested in dark photons much lighter than the $Z$ boson and therefore we need only consider the mixing of the $U(1)_D$ gauge symmetry with $U(1)_\mathrm{QED}$ (rather than with the full hypercharge symmetry $U(1)_Y$ which would be relevant above the Electroweak scale).}, which takes the form:
\begin{equation}
     \mathcal{L}_\mathrm{DAP} \supset \frac{\epsilon}{2} F_{\mu\nu}F^{\prime\mu\nu} \,,
\end{equation}
where $F_{\mu\nu}$ is the usual electromagnetic field strength tensor; $F^\prime_{\mu\nu} = \partial_\mu A^\prime_\nu - \partial_\nu A^\prime_\mu$ is the corresponding field strength tensor for the $U(1)_D$ gauge field $A^\prime$; and $\epsilon$ is the kinetic mixing parameter.

The heavy fermions $\psi$ and $\psi^c$ can carry charge under the Standard Model gauge groups (in particular color charge), as well as under the new $U(1)_\mathrm{PQ}$ and $U(1)_D$ symmetries. They can therefore induce axion-gluon and axion-photon couplings at loop-level:
\begin{equation}
    \mathcal{L}_\mathrm{DAP} \supset \frac{g_{a g g}}{4} a G_{\mu \nu} \tilde{G}^{\mu \nu}+\frac{g_{a \gamma \gamma}}{4} a F_{\mu \nu} \tilde{F}^{\mu \nu}\,,
\end{equation}
where $G_{\mu\nu}$ is the gluon field strength tensor and the tilde denotes the dual field strength $\tilde{X}^{\mu\nu} = \frac{1}{2} \epsilon_{\mu \nu \lambda \rho} X^{\lambda \rho}$.
In addition, interactions involving both the axion $a$ and dark photon $\DP$ may be induced:
\begin{equation}
    \mathcal{L}_\mathrm{DAP} \supset \frac{g_{a\DP\DP}}{4}aF^\prime_{\mu\nu}\Tilde{F}^{\prime\mu\nu}+\frac{g_{a\gamma\DP}}{2}aF_{\mu\nu}\Tilde{F}^{\prime\mu\nu} \,.
\end{equation}
These axion-dark photon interactions arise through loops of heavy fermions, with dark charge $D_\psi$, as well via the mixing between the dark and visible photons. 

The size of the couplings $g_{agg}$, $g_{a\gamma\gamma}$,  $g_{a\gamma\DP}$,  $g_{a\DP\DP}$ will depend on the specific realisation of the axion model, and in particular the charge assignments of the new particles. As pointed out in Refs.~\cite{Kaneta:2016wvf,Kaneta:2017wfh}, the fermions $\psi$ can also contribute to the kinetic mixing $\epsilon$ at loop-level if they carry an SM electric charge $Q_\psi$ and dark charge $D_\psi$. This contribution goes as:
\begin{align}
    \begin{split}
            \Delta \epsilon &\approx \frac{1}{6\pi^2}N_C (e Q_\psi e' D_\psi)\log(\Lambda/m_\psi)\\
            &\sim 0.005 \left(\frac{e'}{e}\right) Q_\psi D_\psi \log(\Lambda/m_\psi) \,,
    \end{split}
    \label{eq:loopmixing}
\end{align}
where $N_C = 3$ the number of colors. The cut-off scale $\Lambda$ is the scale of additional New Physics, where we fix the bare value of $\epsilon$. As we will see later, such a large correction to $\epsilon$ would contradict a number of constraints, including bounds from the overproduction of dark photons in the early Universe. In the absence of fine-tuning in the bare value of $\epsilon$, then, it is wise to suppress these loop-induced contributions. We therefore assume that the new fermions carry no SM electric charge: $Q_\psi = 0$.

With this, the couplings are given by
\begin{align}
    g_{agg}&=\frac{g_S^2}{8\pi^2}\frac{PQ_\Phi}{f_a} \,,
    \label{eq:Gagg}\\
    g_{a\gamma\gamma}&=-\frac{e^2}{8\pi^2}\frac{PQ_\Phi}{f_a}\left(\frac{2}{3}\frac{4+z}{1+z}\right) \,,
    \label{eq:Gapp}\\
    g_{a\gamma\DP}&\simeq\epsilon g_{a\gamma\gamma} \,,
    \label{eq:Gapdp}\\
    g_{a\DP\DP}&\simeq\frac{e^{\prime2}}{8\pi^2}\frac{PQ_\Phi}{f_a}\left(2N_CD_\psi^2\right)+2\epsilon g_{a\gamma\DP} \,,
    \label{eq:Gadpdp}
\end{align}
where $g_S$ is the strong coupling (related to the strong coupling constant as $\alpha_S=g_S^2/4\pi \simeq 0.12$ at high energies). Without loss of generality, we can take the charge of the PQ field to be one ($PQ_\Phi=1$) because the symmetry breaking scale $f_a$ is also proportional to $PQ_\Phi$. Finally, we must fix the dark coupling constant $e^\prime$ and the dark charge of the new heavy fermions $D_\psi$. For concreteness, we will typically consider $e^\prime = 0.1$ and $D_\psi = 3$ unless otherwise stated, such that the new fermions couple to the dark $U(1)_D$ with the same strength as the electron couples to SM electromagnetism: $e^\prime D_\psi \sim e \sim 0.3$. In Appendix~\ref{app:Charges}, we briefly explore how varying the dark charge affects our final results. 

With these choices, the model is then specified by the mass of the dark photon $\mDP$, the mass of the axion $m_a$ (or equivalently the axion decay constant $f_a$, related through Eq.~\eqref{eq:axionmass}), and the kinetic mixing $\epsilon$.


\section{Dark Matter production in the Dark Axion Portal}
\label{sec:Production}

In this section, we discuss how the interactions described above can give rise to a substantial relic abundance of long-lived DM particles. In particular, we will focus on scenarios in which both the axion and the dark photon contribute to the observed DM density in the Universe of $\Omega_\mathrm{DM} h^2 \approx 0.120 \pm 0.001$~\cite{Planck:2018vyg}.\footnote{Here, we use the `small $h$' notation, such that the Hubble parameter today is $H_0 = 100\,h\,\mathrm{km/s}\,\mathrm{Mpc}^{-1}$.}

\subsection{Axion production}
\label{sec:Production:Axions}

Axions can be produced non-thermally through the misalignment mechanism~\cite{Preskill:1982cy,Abbott:1982af,Dine:1982ah}. At the QCD phase transition non-perturbative effects generate a mass for the axion, and the axion field relaxes to the minimum of its potential $V(\theta)$. The oscillation of the axion field around its minimum gives rise to an energy density which acts as cold DM. The initial value of the axion field  in some patch of the Universe $\theta_i$ is chosen at random when the PQ symmetry is broken and, along with the axion decay constant $f_a$, sets the axion abundance~\cite{Marsh:2015xka}.

The relic abundance of axions depends on whether the PQ symmetry is broken before or after the end of inflation. Concretely, if the Hubble rate during inflation $H_I$ exceeds the PQ scale, $H_I/(2\pi) > f_a$, then the PQ symmetry is broken after inflation has ended. When the symmetry is broken, the axion field takes on different initial values $\theta_i \in [0, \pi]$ in different causally disconnected patches of the observable Universe. The energy density in the axion field due to the initial misalignment can be estimated by averaging over different patches, for which $\langle \theta_i^2\rangle = \pi^2/3$~\cite{Turner:1985si}. The breaking of the PQ symmetry also leads to topological defects such as strings and domain walls, the decay of which produces an additional source of axion energy density. State-of-the-art numerical simulations suggest that for axions to make up all the DM in the Universe, their mass must be on the order of $m_a \sim (25.2 \pm 11.0) \,\mu\mathrm{eV}$~\cite{Klaer:2017ond,Buschmann:2019icd} (though the exact contribution of string decay is disputed and may lift the required axion mass to $m_a \gtrsim 500 \,\mu\mathrm{eV}$~\cite{Gorghetto:2020qws,Hindmarsh:2021zkt}). Following Ref.~\cite{Buschmann:2019icd}, the axion density in this scenario is expected to scale as $\Omega_a \sim f_a{}^{1.187}$, allowing us to compute the required $f_a$ (or equivalently $m_a$) also for sub-dominant axion contributions.

Instead, if $H_I/(2\pi) < f_a$, the PQ symmetry is broken before the end of inflation and a single patch is inflated to become the Universe we see today. In this case, the axion energy density is determined by the initial field value of the axion in this patch, which will be randomly distributed in the range $\theta_i \in [0, \pi]$. In addition, the axion field will undergo fluctuations during inflation giving rise to an effective contribution to the misalignment angle of $\sigma_\theta^2 = H_I^2/(2\pi f_a)^2$~\cite{1992PhRvD..45.3394L}.
The axion energy density today is then given by~\cite{Fox:2004kb}:
\begin{equation}
    \Omega_ah^2=0.5\left[\frac{f_a/\mathcal{N}}{10^{12}\textrm{ GeV}}\right]^\frac{7}{6}\left[\theta_i^2+\sigma_\theta^2\right] \mathcal{F}_\mathrm{anh}(\theta_i^2) \,,
    \label{eq:axiondensity}
\end{equation}
where $\mathcal{N}$ is the PQ color anomaly ($\mathcal{N} = 1$ in the KSVZ model we assume), and $\mathcal{F}_\mathrm{anh}(\theta_i^2)$ is a correction coming from the fact that the axion potential is anharmonic for large $\theta_i$~\cite{Turner:1985si,Visinelli:2009zm}.\footnote{Note that $\mathcal{F}_\mathrm{anh}(\theta_i^2) \rightarrow 1$ for $\theta_i \ll 1$, but diverges as $\theta \rightarrow \pi$. In practice, we truncate $\mathcal{F}_\mathrm{anh}$ to have a maximum value of 10, in order to avoid achieving arbitrarily large axion densities for $\theta_i \approx \pi$.} We have assumed that there is no substantial entropy injection which may dilute the axion density at late times.

The backreaction contribution, $\sigma_\theta^2 = H_I^2/(2\pi f_a)^2$, gives rise to a minimum value of the axion density, even if we tune $\theta_i \rightarrow 0$. More problematic, however, is the fact that inflationary fluctuations give rise to isocurvature perturbations in the axion density field~\cite{1983PhLB..126..178A,Kofman:1986wm}, with an amplitude $\propto H_I^2$ which directly probes the energy scale of inflation. Strong constraints from PLANCK on the ratio of isocurvature to adiabatic perturbations requires $H_I \lesssim 10^7 \,\mathrm{GeV}$ for $f_a \lesssim 10^{12}\,\mathrm{GeV}$~\cite{Marsh:2015xka}. The backreaction contribution to the axion density $\sigma_\theta^2$ can therefore safely be neglected, as long as we restrict ourselves to scenarios of low-scale inflation~\cite{German:2001tz,Takahashi:2018tdu,Schmitz:2018nhb}. This subtlety was not previously appreciated in discussions of the Dark Axion Portal.

\begin{figure}
    \centering
    \includegraphics[width=0.47\textwidth]{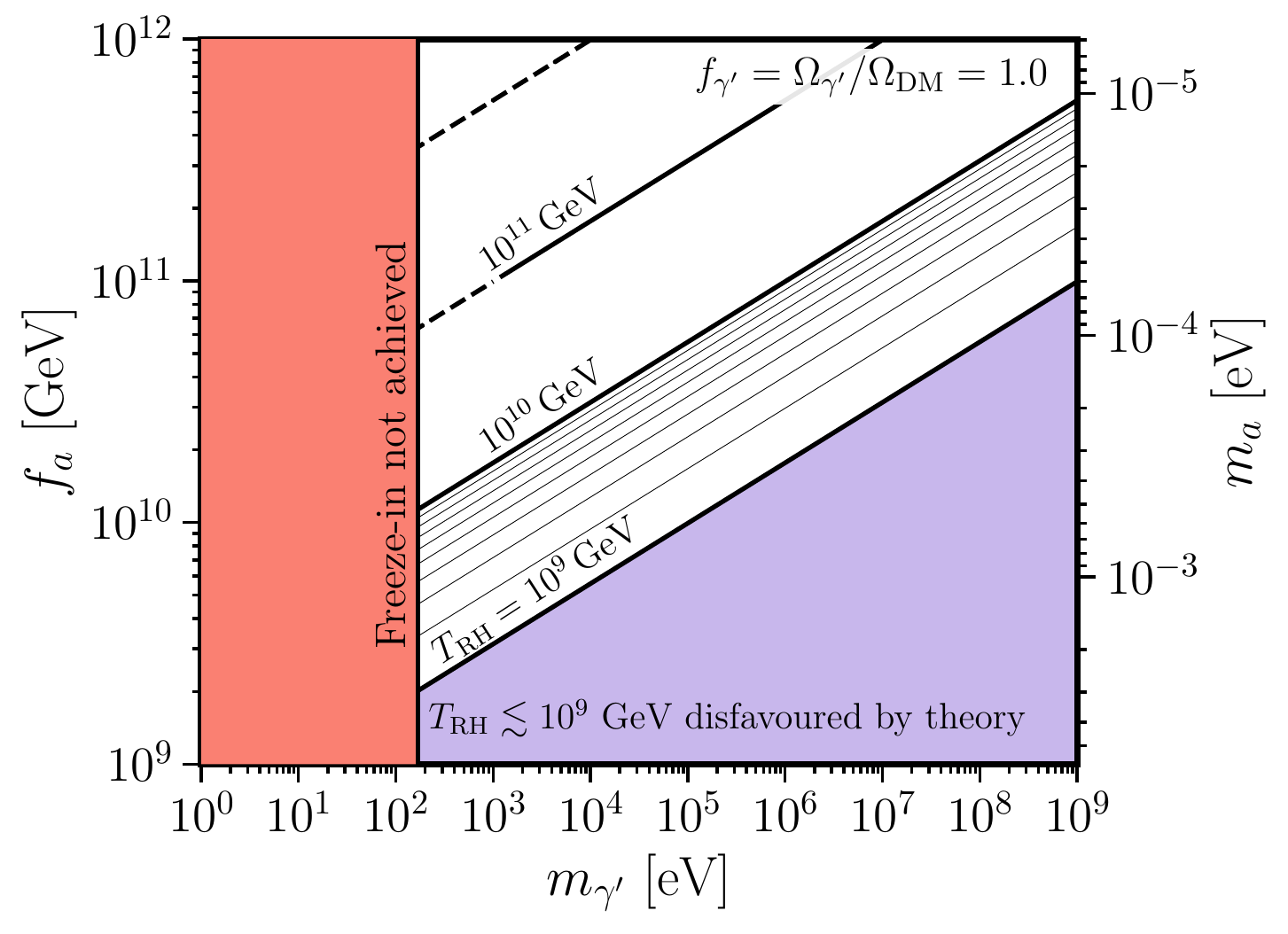}\\
    \includegraphics[width=0.47\textwidth]{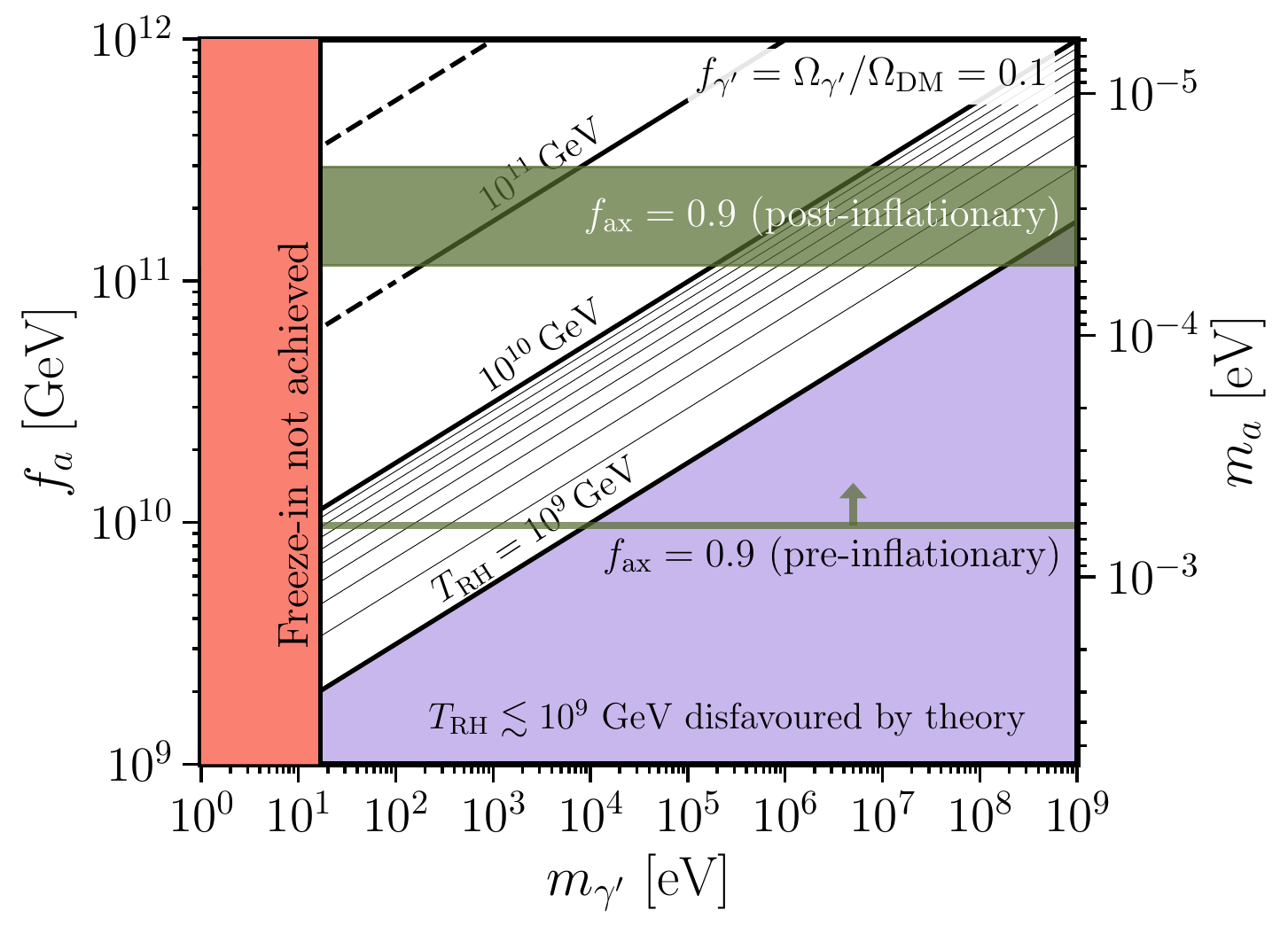}\\
    \includegraphics[width=0.47\textwidth]{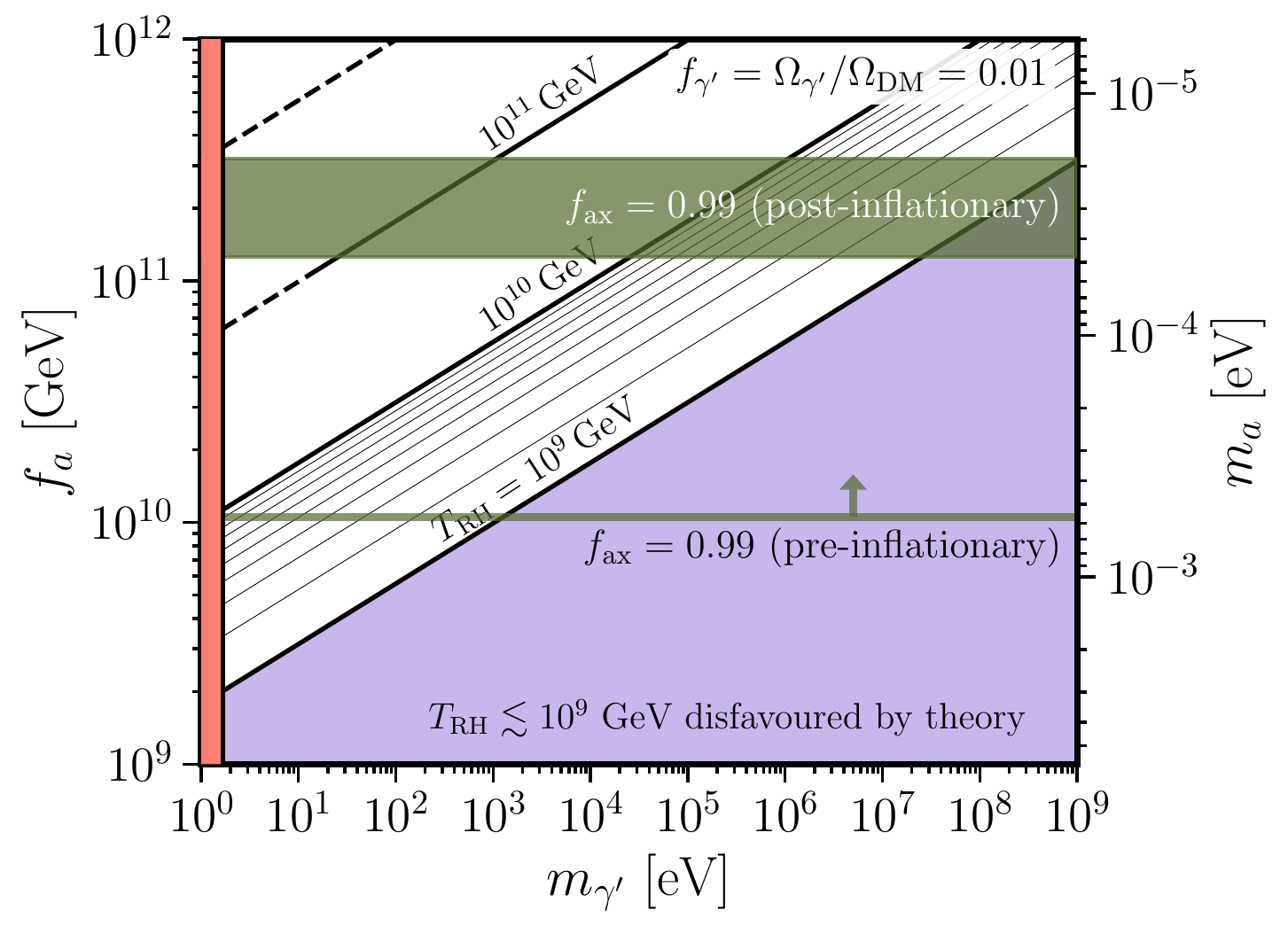}
    \caption{\textbf{Parameter space for dark photon freeze-in through gluon fusion.} White regions show where dark photons constitute 100\% (\textbf{top}), 10\% (\textbf{middle}) and 1\% (\textbf{bottom}) of the total DM density in the Universe (see Sec.~\ref{sec:Production:DarkPhotons:GluonFusion}). Solid black lines show the value of the reheating temperature $T_\mathrm{RH}$ required to achieve this density. For small values of the dark photon mass $m_{\gamma'}$, a large value of the reheating temperature would be required to achieve the desired $f_{\DP}$ causing the dark photon to enter thermal equilibrium in the early Universe and rendering Freeze-in impossible (red regions). Green regions indicate the preferred axion parameters in different production scenarios (see Sec.~\ref{sec:Production:Axions}).}
    \label{fig:FreezeInRegion}
\end{figure}

We therefore consider two possible scenarios for the production of the axion, depending on whether the PQ symmetry was broken \textit{after} or \textit{before} the end of inflation:
\begin{itemize}
	\item \textbf{Post-inflationary, $f_a < H_I/(2\pi)$} - with masses in the range $m_a \sim 20 - 50 \,\mu\mathrm{eV}$, following the simulations of Ref.~\cite{Buschmann:2019icd}. 
	\item \textbf{Pre-inflationary, $f_a > H_I/(2\pi)$} - with masses in the range $m_a \lesssim 600 \,\mu\mathrm{eV}$. In this scenario, we require low-scale inflation $H_I \lesssim 10^7\,\mathrm{GeV}$ so as to avoid the overproduction of isocurvature perturbations by the axion field.
\end{itemize}
We highlight some of these possibilities in Fig.~\ref{fig:FreezeInRegion}, as a function of the axion mass $m_a$ (or equivalently $f_a$). The post-inflationary scenario is shown as a green band (adapting the results of Ref.~\cite{Buschmann:2019icd} to axion fractions $f_\mathrm{ax} = \Omega_a/\Omega_\mathrm{DM} < 1$), while the pre-inflationary scenario is shown as a lower limit on $f_a$.\footnote{Note that we use the notation $f_a$ for the axion decay constant and the notation $f_\mathrm{ax}$ for the DM fraction in axions.} Above this green line, we can fix the axion density to the desired fraction for some value of $\theta_i \lesssim \pi$, following Eq.~\eqref{eq:axiondensity}.

We note that in the Dark Axion Portal, the axion relic abundance can in principle be affected by the coupling to the dark photon. A large abundance of dark photons may be produced through a tachyonic instability when the axion field starts oscillating~\cite{Agrawal:2017eqm,Kitajima:2017peg,Agrawal:2018vin,Co:2018lka}, substantially diluting the axion density. However this mechanism is only effective when $\mDP < m_a$, and requires a coupling $g_{a\DP\DP} f_a \gtrsim \mathcal{O}(10)$~\cite{Agrawal:2017eqm}. For our choice of parameters, we have $g_{a\DP\DP} f_a \sim 10^{-2}$ and we consider the scenario $\mDP > m_a$, meaning that we can safely neglect this process of axion dilution (and the corresponding dark photon production).   

\subsection{Dark Photon production}
\label{sec:Production:DarkPhoton}

We now consider possible mechanisms for the production of dark photons. We first note that a number of proposed mechanisms may not be possible in the Dark Axion Portal Model. For example, inflationary production of dark photons requires high-scale inflation~\cite{Graham:2015rva,Bastero-Gil:2018uel}, which would be at odds with the pre-inflationary scenario for axion production described in Sec.~\ref{sec:Production:Axions}. A misalignment mechanism for the dark photon -- similar to axion misalignment -- is also not possible for a dark photon which gains its mass from a Higgs mechanism~\cite{Nelson:2011sf}, as here. This further emphasises the need to self-consistently study the production of DM; in our case, it is hard to rely on mechanisms outside of the Dark Axion Portal itself in order to produce the desired dark photon density.

\newsavebox{\firstdiagram}
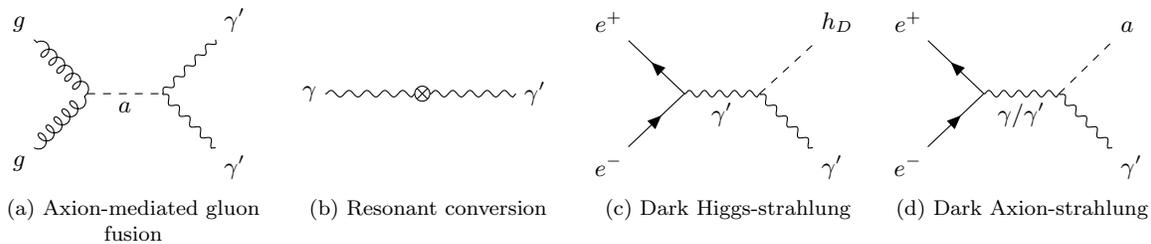
\begin{figure*}[t]
\centering
\savebox{\firstdiagram}{
\begin{tikzpicture}
    \begin{feynman}[small]
    \vertex (b) ;
    \vertex [below left=of b] (a) {\(g\)};
    \vertex [above left=of b] (f1) {\(g\)};
    \vertex [right=of b] (c);
    \vertex [above right=of c] (f2) {\(\gamma^\prime\)};
    \vertex [below right=of c] (f3) {\(\gamma^\prime\)};
    \diagram* {
    (a) -- [gluon] (b) -- [gluon] (f1),
    (b) -- [scalar, edge label'=\(a\)] (c),
    (c) -- [photon] (f2),
    (c) -- [photon] (f3),
    };
    \end{feynman}
    \end{tikzpicture}
}
\subfloat[Axion-mediated gluon fusion]{
    \centering
    \usebox{\firstdiagram}
    \label{fig:GluonFusion}
}
\subfloat[Resonant conversion]{
    \centering
    \raisebox{\dimexpr.5\ht\firstdiagram-.5\height}{
        \begin{tikzpicture}
        \begin{feynman}[small]
        \vertex (a) {\(\gamma\)} ;
        \node [right=1.5cm of a, crossed dot] (b) ;
        \vertex [right=1.5cm of b] (c) {\(\gamma^\prime\)};
        \diagram* {
        (a) -- [photon] (b) -- [photon] (c),
        };
        \end{feynman}
        \end{tikzpicture}
    \label{fig:Feynman:Resonant}
    }
}
\subfloat[Dark Higgs-strahlung]{
    \centering
    \begin{tikzpicture}
    \begin{feynman}[small]
    \vertex (b) ;
    \vertex [below left=of b] (a) {$e^-$};
    \vertex [above left=of b] (f1) {$e^+$};
    \vertex [right=of b] (c);
    \vertex [above right=of c] (f2) {$h_D$};
    \vertex [below right=of c] (f3) {$\gamma^\prime$};
    \diagram* {
    (a) -- [fermion] (b) -- [fermion] (f1),
    (b) -- [photon, edge label'=$\gamma^\prime$] (c),
    (c) -- [scalar] (f2),
    (c) -- [photon] (f3),
    };
    \end{feynman}
    \end{tikzpicture}
    \label{fig:Feynman:DarkHiggs}
}
\subfloat[Dark Axion-strahlung]{
    \centering
    \begin{tikzpicture}
    \begin{feynman}[small]
    \vertex (b) ;
    \vertex [below left=of b] (a) {$e^-$};
    \vertex [above left=of b] (f1) {$e^+$};
    \vertex [right=of b] (c);
    \vertex [above right=of c] (f2) {$a$};
    \vertex [below right=of c] (f3) {$\gamma^\prime$};
    \diagram* {
    (a) -- [fermion] (b) -- [fermion] (f1),
    (b) -- [photon, edge label'=$\gamma/\gamma^\prime$] (c),
    (c) -- [scalar] (f2),
    (c) -- [photon] (f3),
    };
    \end{feynman}
    \end{tikzpicture}
    \label{fig:Feynman:DarkAxion}
}
\caption{\textbf{Processes contributing to the early Universe production of dark photons $\DP$} described in Sec.~\ref{sec:Production:DarkPhoton}. Processes (b) and (c) are only possible for non-zero kinetic mixing $\epsilon \neq 0 $. Process (d) receives contributions from two separate diagrams (mediated by either the visible or dark photon). In the Dark KSVZ model, this process is also negligible for $\epsilon = 0$.}
\label{fig:Feynman}
\end{figure*}

As in Ref.~\cite{Kaneta:2016wvf}, then, we will first discuss the freeze-in production of dark photons via gluon fusion, mediated by the axion, extending the calculation to eV-scale dark photons. We do not consider the Dark Primakoff process described in Ref.~\cite{Kaneta:2017wfh} because we fix the charge of the heavy fermions $Q_\psi = 0$, as described in Sec.~\ref{sec:DarkAxionPortal}. However, we do consider additional channels which open when there is a non-zero kinetic mixing $\epsilon \neq 0$, including resonant photon conversion ($\gamma \rightarrow \gamma^\prime$), Dark Higgs-strahlung and Dark Axion-strahlung. Some of the relevant processes are illustrated in Fig.~\ref{fig:Feynman}.

\subsubsection{Freeze-in through gluon fusion}
\label{sec:Production:DarkPhotons:GluonFusion}

The dark photon-axion coupling in the Dark Axion Portal allows for the gluon fusion process $gg \rightarrow a \rightarrow \DP\DP$. This process is typically slow enough that it does not allow the dark photons to enter into thermal equilibrium with the Standard Model plasma. However, it can be sufficiently large to produce dark photon DM from an initially negligible abundance, through the freeze-in mechanism~\cite{Hall:2009bx}. This process was studied in detail in Refs.~\cite{Kaneta:2016wvf,Kaneta:2017wfh}. Here, we briefly summarise the freeze-in calculation, and extend the analysis to light (eV-scale) dark photons.

The yield of dark photons $Y_{\DP} = n_{\DP}/S$ as a function of temperature $T$ evolves as~\cite{Hall:2009bx}:
\begin{equation}
    \frac{\diff Y_{\gamma'}}{\diff T} = \frac{-\mathcal{C}}{S\,H\,T}\,,
    \label{eq:dYdt}
\end{equation}
with $S$ the entropy density, $n_{\DP}$ the dark photon number density and $H$ the Hubble parameter. The collision term appearing in the  Boltzmann equation $\mathcal{C}$, can be written~\cite{Edsjo:1997bg}:
\begin{align}
\begin{split}
    \mathcal{C} &= \frac{T}{64 \pi^{4}} \int_{0}^{\infty}(\sigma v_\mathrm{rel}) s^{3 / 2} K_{1}\left(\frac{\sqrt{s}}{T}\right) \mathrm{d} s\,,
    \label{eq:CollisionTerm}
\end{split}
\end{align}
where $K_1$ is the modified Bessel function of the second kind, $s$ is the centre-of-mass energy, and the annihilation cross-section for the gluon-fusion process in Fig.~\ref{fig:GluonFusion} is~\cite{Kaneta:2016wvf}
\begin{equation}
\sigma_{gg\rightarrow\DP\DP} v \sim 4g_{agg}^2g_{a\DP\DP}^2s\,.
\end{equation}
With this, the collision term becomes:
\begin{equation}
    \mathcal{C} = \frac{48}{\pi^4}g_{agg}^2g_{a\DP\DP}^2T^8\,.
\end{equation}
The yield today is obtained by integrating from the reheating temperature $T_\mathrm{RH}$ to $T = 0$, assuming an initially negligible number density of dark photons:
\begin{equation}
    Y_{\DP}(T=0) =\int_{0}^{T_\mathrm{RH}}\frac{\mathcal{C}}{S\,H\,T}\dd{T}\,.
    \label{eq:Yield2}
\end{equation}
The entropy density and Hubble parameter during radiation domination can be written as~\cite{Baumann:2018muz}:
\begin{equation}
    S = \frac{2\pi^2}{45}g_{\star S}(T) T^3\,; \quad H^2 = \frac{\pi^2}{90}g_\star(T) \frac{T^4}{M_\mathrm{pl}^2}\,,
\end{equation}
with the Planck mass $M_\mathrm{pl} \approx 2.4 \times 10^{18}\,\mathrm{GeV}$, and $g_{\star}$ ($g_{\star,S}$) the effective number of relativistic degrees of freedom (in entropy). The cross section grows rapidly with $T$, meaning that the production of dark photons is dominated by temperatures close to $T_\mathrm{RH}$. We therefore fix $g_\star = g_{\star,S} = 106.75$ to obtain the yield as: 
\begin{equation}
    Y_{\DP}(T=0)
    \approx 10^{-3} g_{agg}^2 g_{a\DP\DP}^2  T_\mathrm{RH}^3 M_{\mathrm{pl}}\,.
    \label{eq:completeYield}
\end{equation}
The energy density today is then:
\begin{equation}
\Omega_{\DP, 0}\,h^2 = \frac{1}{\rho_{c,0} h^{-2}} m_{\DP} S_0 Y_{\DP}
\end{equation}
where $S_0 \approx 2889.2 \,\mathrm{cm}^{-3}$ is the present day entropy density and $\rho_{c,0}h^{-2} = 1.05368 \times 10^{-5} \,\mathrm{GeV}\,\mathrm{cm}^{-3}$ is the critical density. Finally, the fraction of DM in dark photons $f_{\DP} = \Omega_{\DP, 0}/\Omega_{\mathrm{DM}, 0}$ can be written:
\begin{equation}
    f_{\DP} \approx \left(\frac{e' D_\psi}{0.3}\right)^4 \left(\frac{\mDP}{10\,\mathrm{eV}}\right)\left(\frac{10^9\,\mathrm{GeV}}{f_a}\right)^4\,\left(\frac{T_\mathrm{RH}}{10^9\,\mathrm{GeV}}\right)^3\,.
    \label{eq:DP_fraction}
\end{equation}

In Fig.~\ref{fig:FreezeInRegion}, we show the regions of the parameter space $(\mDP, f_a)$ where it is possible to achieve a dark photon fraction of 100\% (top), 10\% (middle) and 1\% (bottom), with black lines indicating the required value of $T_\mathrm{RH}$ in each case. The reheating temperature is \textit{a priori} unknown, and could be as low as a few MeV, while still giving rise to successful Big Bang Nucleosynthesis~\cite{Hannestad:2004px}. However, there are theory motivations for a larger value of $T_\mathrm{RH}$. For example, the successful generation of the lepton asymmetry in the Universe through thermal leptogenesis typically requires a reheat temperature greater than $10^8 - 10^{10}\,\mathrm{GeV}$~\cite{Davidson:2002qv,Giudice:2003jh,Buchmuller:2005eh,Hahn-Woernle:2008tsk}. We therefore shade the region $T_\mathrm{RH} \lesssim 10^9\,\mathrm{GeV}$ in purple in Fig.~\ref{fig:FreezeInRegion}, though we note that this is not a hard bound and leptogenesis constraints can be evaded~\cite{Boubekeur:2002jn,Domcke:2020quw}.

As pointed out in Ref.~\cite{Kaneta:2016wvf}, the reheating temperature is constrained by requiring that the dark photon does not enter thermal equilibrium with the Standard Model plasma. This requirement can be phrased as $\mathcal{C}/n_{\DP}^{\mathrm{eq}} < H$, where $n_{\DP}^{\mathrm{eq}} = (3\zeta(3)/\pi^2)T^3$ is the equilibrium dark photon number density. Here, we point out that for a fixed $f_{\DP}$, this maximum reheating temperature can be recast as a minimum dark photon mass as:
\begin{equation}
    m_{\DP} \gtrsim f_{\DP} \times 168 \,\mathrm{eV}\,.
    \label{eq:m_min}
\end{equation}
Below this mass, the dark photon abundance would be set by freeze-out, rather than the freeze-in mechanism we summarise here. We show this bound as a red shaded region in Fig.~\ref{fig:FreezeInRegion}.

Crucially, if we hope to obtain dark photons with masses in the range $1 - 100\,\mathrm{eV}$, which may be detectable at direct detection experiments, we require $f_{\DP} \lesssim 10\%$ (see the lower two panels of Fig.~\ref{fig:FreezeInRegion}). In these scenarios, it is possible to produce enough axions to constitute the remaining $f_\mathrm{ax} > 90\%$ of the DM abundance, in both the pre-inflationary and post-inflationary scenarios. 

For the values of the axion decay constant $f_a$ indicated by these axion production mechanisms, Fig.~\ref{fig:FreezeInRegion} suggests that to achieve the desired dark photon abundance in the mass range of interest would require a reheating temperature of $T_\mathrm{RH} \gtrsim 10^9\,\mathrm{GeV}$. For very large $T_\mathrm{RH}$, we may find that $T_\mathrm{RH} > f_a$, for which we plot the $T_\mathrm{RH}$ contours in Fig.~\ref{fig:FreezeInRegion} as dashed lines instead of solid. When this occurs, the PQ symmetry will be restored during re-heating, regardless of whether it was broken during inflation. In this case, the axion density will be set by the post-inflationary mechanism. As we will see later, this sets an important constraint on the possible re-heating temperature (and therefore axion mass) in the pre-inflationary scenario.

\subsubsection{Resonant Conversion}
\label{sec:Production:DarkPhoton:Conversion}

We now consider the production of dark photons through alternative channels, which open once we allow for a non-zero value of the kinetic mixing $\epsilon$.

There are several important mass regimes where different production mechanisms become relevant. If the dark photon mass is greater than twice the electron mass, the coalescence channel $e^+ + e^- \rightarrow \DP$ is open and typically dominates the production. Given that we focus in this work on light dark photons $m_{\gamma\prime}< 1\, \mathrm{MeV}$, the coalescence channel is kinetically forbidden, and the dominant channel is the resonant conversion of photons into dark photons $\gamma \rightarrow \DP$, as illustrated in Fig.~\ref{fig:Feynman:Resonant}.\footnote{In principle, there is also a Compton-like channel $\gamma e \rightarrow \DP e$, though this is typically suppressed with respect to resonant conversion by higher powers of $\alpha = e^2/(4\pi)$, the electromagnetic fine-structure constant~\cite{Redondo:2008ec}.}

The small kinetic mixing leads to photon oscillations in vacuum, with a small probability $\sim \epsilon^2$ of oscillation into a dark photon~\cite{Redondo:2008ec}. In the early Universe, however, the photons are in a thermal plasma, so matter effects should be taken into account~\cite{Dvorkin:2019zdi,Dvorkin:2020xga}. Compton scattering with electrons gives rise to an effective mass (or plasma mass) for the photon $m_\gamma$. In this environment, the mixing probability becomes~\cite{Redondo:2008aa}:
\begin{equation}
    \epsilon_\mathrm{eff}^2(\omega,T)\simeq\epsilon^2\frac{m_{\gamma\prime}^4}{\left(m_{\gamma\prime}^2-m_{\gamma}^2\right)^2+(\omega D)^2}\,,
    \label{eq:mixingparameter}
\end{equation}
where $\omega$ is the photon energy, and $D$ is a damping factor (see Appendix A of Ref.~\cite{Redondo:2008ec} for further details). When $\mDP \sim m_\gamma$, the mixing is resonantly enhanced, which for light dark photons $\mDP \in [1, 10^5]\,\mathrm{eV}$ occurs at a temperature close to $T_\mathrm{res} \sim 0.2 m_e$, when the electrons are non-relativistic.

Following Ref.~\cite{Redondo:2008ec}, assuming that the dark photon yield is initially negligible, the yield at late times is given by:
\begin{align}
\begin{split}
    Y_{\DP}&\simeq\epsilon^2\frac{\pi\zeta(2)}{\zeta(3)}\frac{m_{\DP}^2Y_\gamma}{H\,T\,j(T)}\frac{\dd \ln{s}}{\dd \ln{T^3}}\bigg\rvert_{T=T_\mathrm{res}}\\
    &\simeq 3.4 \times 10^{11} \epsilon^2 \left(\frac{\mDP}{\mathrm{eV}}\right)^2\,,
\end{split}
\end{align}
where $j(T) \equiv (T/m_\gamma^2)(\dd m_\gamma^2/\dd T) = 3$ for non-relativistic electrons and we evaluate the photon yield $Y_{\gamma}$ at the resonance temperature. Again, converting this yield into a dark photon fraction in Dark Matter, we obtain:\footnote{Note that this result differs from the final result for the dark photon abundance in Ref.~\cite{Redondo:2008ec}, as we generally consider light dark photons, $\mDP \ll m_e$ for which $T_\mathrm{res} \sim 0.2 m_e$ and the electrons are non-relativistic close to the resonance.}
\begin{equation}
    f_{\DP} \simeq 7.7 \times 10^{-4}
    \left(\frac{\mDP}{10\,\mathrm{eV}}\right)^3
    \left(\frac{\epsilon}{10^{-9}}\right)^2 \,.
\end{equation}

In Fig.~\ref{fig:KineticMixingConstrained}, we show the required value of the kinetic mixing $\epsilon$ to obtain $f_{\DP} = 100\%$ and $f_{\DP} = 1\%$ (white dotted lines, labelled `$\gamma \rightarrow \DP$').  We can compare this to a range of known constraints on the dark photon mass and mixing parameter~\cite{Caputo:2021eaa}. We see that for dark photons lighter than $10^4\,\mathrm{eV}$, such large values of $\epsilon$ are excluded by several independent probes, including searches for dark photon emission from the Sun by the CAST~\cite{CAST:2017uph} and SHIPS~\cite{Schwarz:2015lqa} experiments as well as constraints from anomalous stellar cooling (see Sec.~\ref{sec:Production:OtherConstraints}). In the range $\mDP \in [10^4,10^5]\,\mathrm{eV}$, resonant conversion may contribute appreciably to the relic abundance even with kinetic mixings as small as $\epsilon \sim 10^{-14}$, roughly within the sensitivity range of current high-mass searches such as XENON~\cite{XENON:2019gfn,XENON:2020rca}. However, outside of this region, the contribution from resonant conversion is negligible.

We note here that including a non-zero EM charge for the heavy fermions $\psi$ would induce a substantial correction to the kinetic mixing of order $10^{-3}$, as given in Eq.~\eqref{eq:loopmixing}. Such a large value of $\epsilon$ would lead to overproduction of dark photons through resonant conversion (as well as through the Dark Higgs-strahlung process we describe next), further justifying our choice to fix $Q_\psi = 0$.

\subsubsection{Dark Higgs-strahlung}
\label{sec:Production:DarkPhoton:DarkHiggsstrahlung}

Because the dark photon gains its mass through a Higgs mechanism, we must also consider production through processes with Dark Higgs particles $h_D$ in the final state~\cite{Pospelov:2008jk,Batell:2009yf}. Such a `Dark Higgs-strahlung' process is illustrated in Fig.~\ref{fig:Feynman:DarkHiggs} and is only possible with a non-zero kinetic mixing, which allows the production of the $\DP$ mediator from $e^\pm$ annihilation.

\begin{figure}[t]
    \centering
    \includegraphics[width=0.48\textwidth]{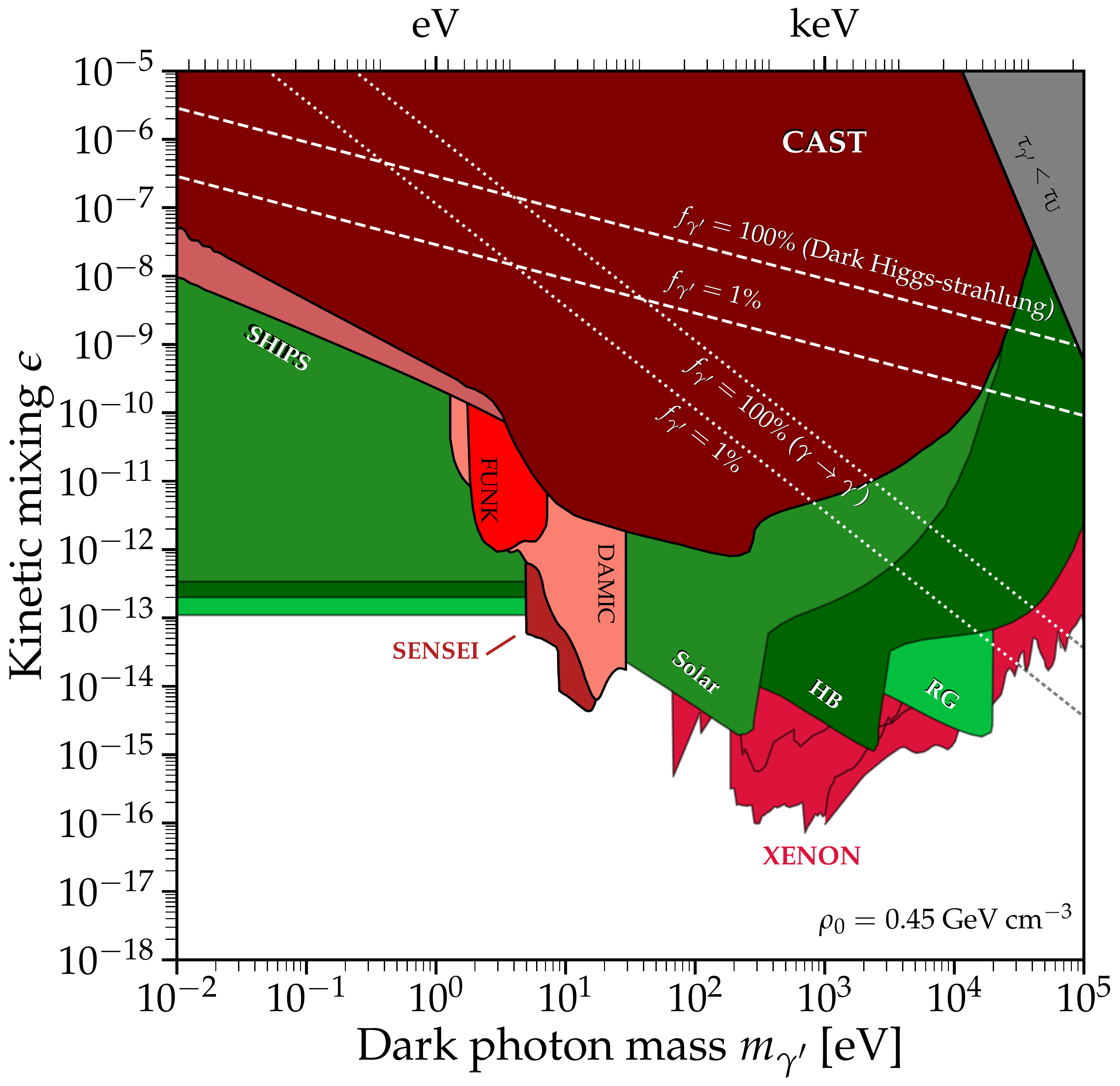}
    \caption{\textbf{Dark Photon production through kinetic mixing $\epsilon$ and associated constraints.} Dotted white lines show the required kinetic mixing to produce $f_{\gamma'} = 1\%$ and  $f_{\gamma'} = 100\%$ from resonant $\gamma \rightarrow \gamma'$ conversion (Sec.~\ref{sec:Production:DarkPhoton:Conversion}), while dashed lines show the required mixing for Dark Higgs-strahlung production (Sec.~\ref{sec:Production:DarkPhoton:DarkHiggsstrahlung}). In the mass range of interest, such large values of $\epsilon$ are strongly constrained by direct and indirect searches; dark photon production through kinetic mixing will contribute negligibly to the present day abundance of eV-scale dark photons. Constraints from stellar cooling (``Solar'', ``HB'', ``RG'') are valid for a Higgsed dark photon, with $e' = 0.1$. Constraints adapted from Ref.~\cite{Caputo:2021eaa}.}
    \label{fig:KineticMixingConstrained}
\end{figure}

In the limit $m_{h_D}, \mDP \ll m_e$, the cross section for Dark Higgstrahlung can be written~\cite{Batell:2009yf}:
\begin{equation}
    \sigma_{e^+e^- \rightarrow h_D \DP} \simeq \frac{\pi \alpha \alpha' \epsilon^2}{3s}\,,
\end{equation}
where $\alpha^\prime = (e')^2/(4\pi)$ is the Dark fine-structure constant ($\alpha^\prime \approx 8 \times 10^{-4}$ for $e^\prime  = 0.1$).
The collision term for production through Dark Higgs-strahlung can be written:
\begin{align}
\begin{split}
    \mathcal{C} &= \frac{3}{2}\frac{T}{64 \pi^{4}} \int_{4 m_e^2}^{\infty}(\sigma v_\mathrm{rel}) s^{3 / 2} K_{1}\left(\frac{\sqrt{s}}{T}\right) \mathrm{d} s\,,
    \label{eq:Collision3}
\end{split}
\end{align}
which differs from Eq.~\eqref{eq:CollisionTerm} in the restriction $s > 4 m_e^2$, taking into account the finite electron mass. The factor of $(3/2)$ appears because we assume that each Dark Higgs produced will subsequently decay into a pair of dark photons (producing 3 dark photons for each $e^\pm$ pair). 
The relative velocity is~\cite{Cannoni:2015wba}:
\begin{align}
\begin{split}
        v_\mathrm{rel} 
        &= \frac{\sqrt{s(s-4 m_e^2)}}{s - 2 m_e^2}\,.
\end{split}
\end{align}
From this, we can write:\footnote{In going from the first line to the second, we neglect the term in square brackets and use the result that $\int_z^\infty x^2 K_1(x)\,\diff x = z^2 K_2(z)$. In the final line, we use that for small $z$, $K_2(z) \approx 2/z^2$, rapidly dropping to zero for $z\gtrsim 1$. Including the term in square brackets (which becomes relevant for $T < m_e$) and performing the analysis numerically increases the yield by about 8\%.}
\begin{align}
\begin{split}
    \mathcal{C} & = \frac{ \alpha \alpha' \epsilon^2}{64 \pi^3} T^4 \int_{2m_e/T}^\infty \left[\frac{\sqrt{1 - \frac{4 m_e^2}{x^2T^2}}}{1 - \frac{2 m_e^2}{x^2 T^2}}\right] x^2 K_1(x)\,\diff x\\
    &\approx \frac{\alpha \alpha' \epsilon^2}{64 \pi^3} T^4 \left(\frac{2m_e}{T}\right)^2 K_2\left(\frac{2 m_e}{T}\right)\\
    &\approx \frac{ \alpha \alpha' \epsilon^2}{32 \pi^3} T^4 \quad\text{for}\quad T \gtrsim m_e\,.
\end{split}
\end{align}
Following the same logic as in Eq.~\eqref{eq:Yield2}, the dark photon yield (starting from zero initial abundance) is given by:
\begin{equation}
\label{eq:YieldIntegral}
    Y_{\gamma'} \approx \frac{135 \sqrt{10} \alpha \alpha' \epsilon^2}{64 \pi^6}M_\mathrm{pl} \int_{m_e}^{T_\mathrm{RH}}  \frac{\diff T}{g_{\star S}(T)\sqrt{g_{\star}(T)}T^2} \,. 
\end{equation}
The yield will be dominated by temperatures close to $T \gtrsim m_e$, so in this case we fix $g_{\star S} = g_\star = 10.75$ (corresponding to relativistic electrons, neutrinos and photons). With this, we obtain:
\begin{equation}
    Y_{\gamma'} \approx \frac{135 \sqrt{10} \alpha \alpha' \epsilon^2}{64 \pi^6 g_{\star}^{3/2}} \frac{M_\mathrm{pl}}{m_e} \approx 0.93 \alpha \alpha' \left(\frac{\epsilon}{10^{-9}}\right)^2\,. 
\end{equation}
This corresponds to an energy density today of:
\begin{align}
\Omega_{\gamma', 0}\,h^2 &= 2.55 \alpha \alpha' \left(\frac{m_{\gamma'}}{10\,\mathrm{eV}}\right) \left(\frac{\epsilon}{10^{-9}}\right)^2\,,
\end{align}
or equivalently:
\begin{equation}
    f_{\DP} \simeq 1.2 \times 10^{-4} \left(\frac{e^\prime}{0.1}\right)^2\left(\frac{m_{\gamma'}}{10\,\mathrm{eV}}\right) \left(\frac{\epsilon}{10^{-9}}\right)^2\,.
\end{equation}
Requiring that the dark photons make up no more than 100\% of the energy density of the Dark Matter, $f_{\DP} < 1$, we thus require:
\begin{align}
\epsilon \lesssim 9.1 \times 10^{-8} \left(\frac{0.1}{e^\prime}\right) \left(\frac{m_{\gamma'}}{10\,\mathrm{eV}}\right)^{-1/2}\,, 
\end{align}
a value which is already well constrained (see Fig.~\ref{fig:KineticMixingConstrained}) and substantially larger than the kinetic mixing values we are aiming to detect. Dark Higgs-strahlung therefore does not threaten to overclose the Universe with dark photons.

\subsubsection{Dark Axion-strahlung}

The final production process we consider is Dark Axion-strahlung, as illustrated in Fig.~\ref{fig:Feynman:DarkAxion}. This process receives contributions from two separate diagrams: one mediated by the dark photon, with amplitude proportional to $\epsilon g_{a\DP\DP}$, and another mediated by the visible photon, with amplitude proportional to $g_{a\gamma\DP}$. In the Dark KSVZ model we consider here, $g_{a\gamma\DP} \simeq \epsilon g_{a\gamma\gamma}$, meaning that the two diagrams may give comparable contributions. Similar processes have been considered previously in the context of collider searches~\cite{Dolan:2017osp,deNiverville:2018hrc,Biswas:2019lcp,Wang:2021uyb}.

With the help of FeynCalc~\cite{Mertig:1990an,Shtabovenko:2016sxi,Shtabovenko:2020gxv}, we find that the total cross section for the Dark Axion-strahlung process is:
\begin{equation}
    \sigma_{e^+e^- \rightarrow \DP a} \simeq \frac{\alpha  \left(2 m_e^2+s\right) \left(\epsilon  g_{a\DP\DP}+g_{a\gamma \gamma^\prime}\right)^2}{24 \sqrt{s^2-4 m_e^2s}}\,,
\end{equation}
where we have taken the limit $\mDP \rightarrow 0$.\footnote{The visible photon-mediated part of this process was considered in Ref.~\cite{deNiverville:2018hrc} in the context of axion and dark photon production at $B$ factories. We have verified that our cross section calculations match those in Ref.~\cite{deNiverville:2018hrc} in the limit $m_e \rightarrow 0$.} The outgoing axion and dark photon are both stable and contribute to the DM density, though given that $m_{\DP} \gg m_a$, the contribution to the axion density \textit{per $e^\pm$ annihilation} will be negligible compared to the dark photon density.

Calculation of the collision term proceeds as in Eq.~\eqref{eq:Collision3}, replacing the prefactor $(3/2)$ by $(1/2)$, as only one dark photon is produced per collision. The collision term grows rapidly with $T$, so we neglect the precise behaviour close to the threshold $T \sim 2 m_e$ and write:
\begin{equation}
    \mathcal{C} = \frac{\alpha  \left(\epsilon  g_{a\DP\DP}+g_{a\gamma \gamma^\prime}\right)^2 T^6}{96 \pi^4}\quad \text{for } T \gtrsim m_e\,,
\end{equation}
and zero otherwise. Evaluating the yield as we have done previously, we find:
\begin{equation}
    Y_{\DP} \approx 5 \times 10^{-9} \, \left(\epsilon  g_{a\DP\DP}+g_{a\gamma \gamma^\prime}\right)^2 T_\mathrm{RH}^2 M_\mathrm{pl}\,.
\end{equation}
Note that the term in $\epsilon g_{a\DP\DP}$ is roughly twice as large as the term in $g_{a\gamma \gamma^\prime}$ but with the opposite sign, leading to a mild cancellation (assuming $D_\psi = 3$ and $e^\prime = 0.1$). The present day dark photon density contributed by Dark Axion-strahlung is then:
\begin{equation}
    f_{\DP} \approx 10^{-3} \, \epsilon^2\left(\frac{\mDP}{10\,\mathrm{eV}}\right)\left(\frac{10^9\,\mathrm{GeV}}{f_a}\right)^2\,\left(\frac{T_\mathrm{RH}}{10^9\,\mathrm{GeV}}\right)\,.
\end{equation}
Given that current constraints on $\epsilon$ lie in the range $\epsilon \lesssim 10^{-13}$, we find that this process produces a negligible population of dark photons, unless the axion decay constant $f_a$ is unfeasibly small or the reheat temperature $T_\mathrm{RH}$ is unfeasibly large.

\subsection{Other constraints}
\label{sec:Production:OtherConstraints}

The results of the previous sections suggest that the dominant production mechanism for eV-scale dark photons in the Dark Axion Portal will be their thermal production at high temperature, close to $T_\mathrm{RH}$ through gluon-fusion. In such a `UV freeze-in' scenario, the light dark photons will be highly relativistic at production, with a large free-streaming length. This hot Dark Matter could potentially erase structure on small scales, which would be observable in (for example) galaxy number counts~\cite{Dekker:2021scf}, Lyman-$\alpha$ observations~\cite{Garzilli:2019qki} and measurements of the Cosmic Microwave Background~\cite{Hannestad:2010yi}. 

The precise constraints depend on the phase space distribution of the DM, which in turn depends on the production mechanism~\cite{Bae:2017dpt,Dvorkin:2019zdi,Dvorkin:2020xga}. However, the general consensus is that for particles making up all of the Dark Matter, masses greater than $m_\mathrm{DM} \gtrsim 1 - 10 \,\mathrm{keV}$ are required to avoid constraints from structure formation, somewhat heavier than the eV-scale particles we focus on here. However, in the Dark Axion model, axions produced through misalignment (Sec.~\ref{sec:Production:Axions}) would act as cold DM and as long as axions make up a sufficiently large fraction of DM, they should allow for the successful formation of the observed small-scale structure. Analysis of such `mixed' DM models suggest that the allowed fraction of non-cold DM is between $f_\mathrm{ncdm} \sim 10^{-2} - 10^{-1}$, if the non-cold component has a mass in the range $m_\mathrm{ncdm} \sim 1 - 10^{2}\,\mathrm{eV}$, as we study here~\cite{Diamanti:2017xfo}. We therefore focus on scenarios in which the dark photon is sub-dominant, though not negligible, $f_{\DP} < 10\%$. Note that the constraint on successful freeze-in for the mass range of interest, given in Eq.~\eqref{eq:m_min}, points to similar values of $f_{\DP}$.

Light new particles can also contribute to stellar cooling, as they may be produced in dense stellar environments and escape without further interactions, carrying away energy in the process. Constraints from the cooling of Horizontal Branch (HB) and Red Giant (RG) stars require  $\epsilon \lesssim 10^{-13}\,(e/e')$ for dark photons lighter than the keV-scale~\cite{Ahlers:2008qc,An:2013yua,An:2020bxd} which gain their mass through a Dark Higgs mechanism, as we consider here (see Fig.~\ref{fig:KineticMixingConstrained}). Additional stellar cooling may also be produced in the Dark Axion portal through plasmon decays of the form $\gamma \rightarrow a + \DP$, in which both axions and dark photons carry away energy. This constrains the coupling $g_{a\gamma\DP} \lesssim 6 \times 10^{-9} \,\mathrm{GeV}^{-1}$~\cite{Kalashev:2018bra}. In our Dark KSVZ model, this translates into a constraint on the kinetic mixing of $\epsilon \lesssim 3\times 10^3 (f_a/ 10^9 \,\mathrm{GeV})$, which is much weaker than the other constraints on $\epsilon$ we consider.

When kinetic mixing between the visible and dark photons is allowed, the dark photon is allowed to decay to three photons, with decay width~\cite{Pospelov:2008jk}:
\begin{equation}
    \Gamma(\DP \rightarrow 3\gamma) \approx \left(5\times 10^{-8}\right)\epsilon^2\left(\frac{e^2}{4\pi^2}\right)^4\left(\frac{m_{\DP}^9}{m_e^8}\right) \,.
\end{equation}
Such decays impose a constraint on the maximum mass of the dark photon, by requiring that the lifetime is larger than the age of the Universe $t_\mathrm{univ} = (13.79 \pm 0.02) \times 10^9$ years~\cite{Planck:2018vyg}. For shorter lifetimes, the dark photon density is depleted by the present day, making relic dark photons undetectable at direct detection experiments.
This constraint becomes $\epsilon \lesssim 5.8 \times 10^8 \,(\mDP/(10 \,\mathrm{eV}))^{-9/2}$. We plot this constraint in Fig.~\ref{fig:KineticMixingConstrained}, where it appears only for large values of $\mDP$, in regions of parameter space which are already ruled out. We therefore need not worry about this decay channel for eV-scale dark photons.

The Dark Axion Portal also opens other decay modes. One is a dark photon decaying to an electron-positron pair, but this decay is kinematically forbidden at the mass scale we are considering. Another is dark photon decay to an axion and a photon:
\begin{equation}
    \Gamma(\DP\rightarrow\gamma a)=\frac{g_{a\gamma\gamma\prime}^2}{96\pi}m_{\gamma\prime}^3\left[1-\frac{m_a^2}{m_{\gamma\prime}^2}\right]^3 \,.
\end{equation}
In the Dark KSVZ model we consider, the decay rate for this channel scales as $\Gamma \sim \epsilon^2 \mDP^3/f_a^2$ and is therefore negligible for the parameter values of interest here. If we allow the heavy fermions to carry electromagnetic charge $Q_\psi \neq 0$, however, the decay channel $\DP\rightarrow\gamma a$ may become relevant~\cite{Kaneta:2017wfh}.

Axions are also subject to decay. Assuming that $m_a < \mDP$, the only open decay channel for the axion is into two photons, with rate~\cite{Kaneta:2016wvf}:
\begin{equation}
    \Gamma(a \rightarrow \gamma \gamma) = \frac{g_{a\gamma\gamma}^2}{64\pi} m_a^3 \approx 2.5 \times 10^{-8} \frac{m_a^3}{f_a^2}\,.
\end{equation}
The large hierarchy between the axion mass and the axion decay constant guarantees that we can also neglect axion decays.

\section{Direct detection prospects for Axions and Dark Photons}
\label{sec:Detection}

As we have seen, the Dark Axion Portal provides a mechanism for the production of stable DM candidates in the form of axions and dark photons. We now discuss how these relic particles may be detected in the present day in direct detection experiments.

\subsection{Axion detection}

There are many ways to detect axions and ALPs~\cite{Sikivie:1983ip}. These may rely on the detection of axions produced in the Sun, or at colliders, or on the population of relic DM axions (for a review of the different techniques see Ref.~\cite{Irastorza:2018dyq}). Here, we focus on the latter scenario, in which `axion haloscopes' aim to detect DM axions in the halo of the Milky Way Galaxy. This approach aims to detect the photons that result from the inverse Primakoff conversion of axions to photons in the presence of a strong magnetic field. The resulting electromagnetic signal has a central frequency  determined by the axion mass $\nu = m_a/(2\pi)$ and a line width given by the kinetic energy of the axions in our Galaxy at the location of the Solar system~\cite{OHare:2017yze}. 
The non-detection of a signal allows us to impose limits on the quantity $g_{a\gamma\gamma}\sqrt{\rho_a/\rho_0}$, where $\rho_a$ is the local axion density and $\rho_0=0.45\,\mathrm{GeV} \,\mathrm{cm}^{-3}$ is the total local DM density~\cite{Read:2014qva}.
In the case that the axions are not the only component of dark matter, then the sensitivity to $g_{a\gamma\gamma}$ is reduced as $1/\sqrt{f_\mathrm{ax}}$.

\begin{figure}[t]
    \centering
    \includegraphics[width=0.48\textwidth]{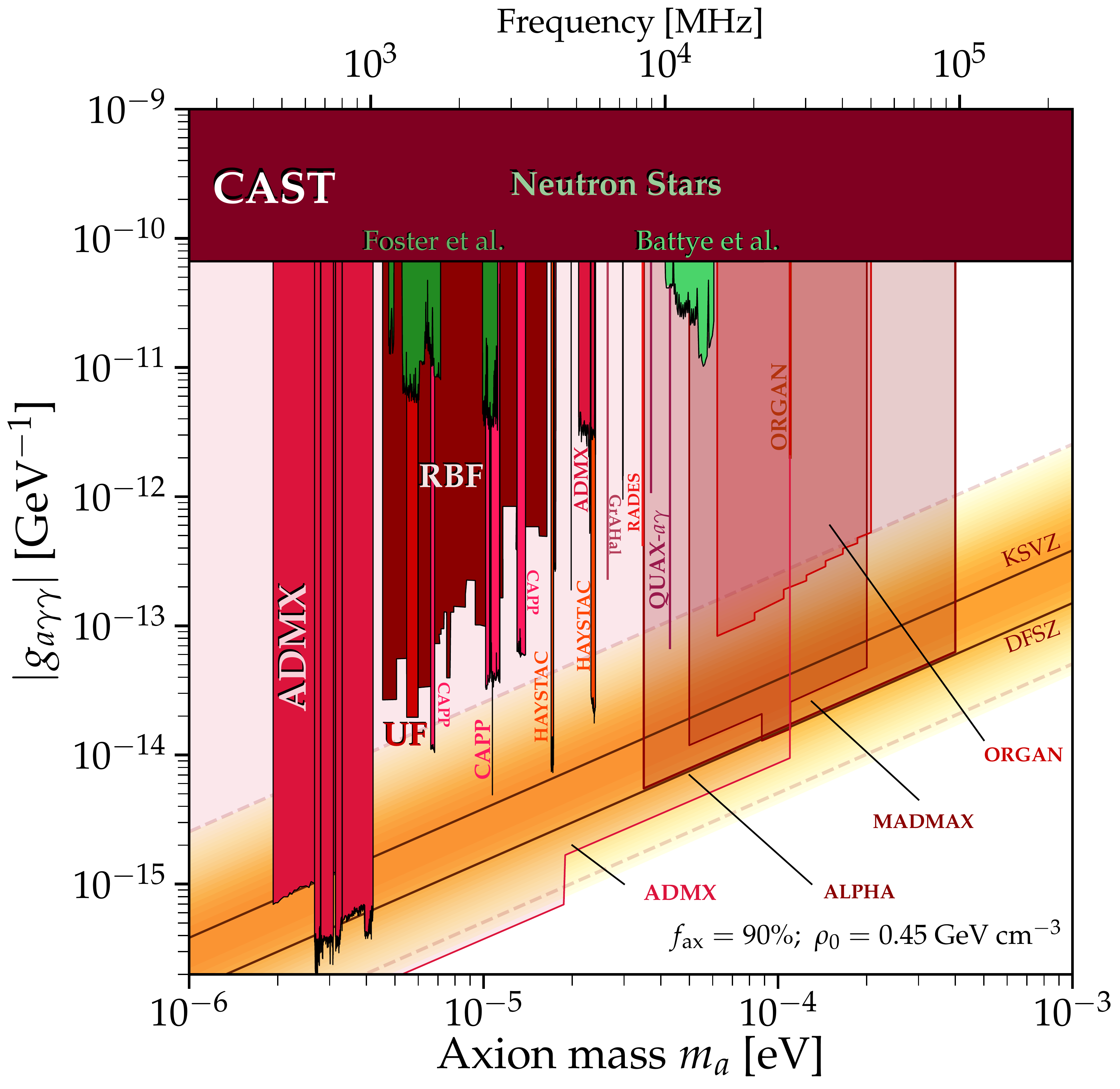}
    \caption{\textbf{Projected constraints on Dark Matter axions}, assuming $f_\mathrm{ax} = 90\%$. Vertical red regions show current constraints from axion haloscopes, with transparent red regions showing projections for proposed haloscope experiments. The solid red region at large $g_{a\gamma\gamma}$ is excluded by the CAST axion helioscope~\cite{CAST:2017uph}, while green regions are excluded by searches for emission from Neutron Stars~\cite{Foster:2020pgt,Battye:2021yue}. 
    Planned haloscope experiments should be able to probe the QCD axion (orange band) over the entire mass range $m_a \sim 0.7 - 400\,\mu\mathrm{eV}$. Constraints adapted from Ref.~\cite{axionlimits}.}
    \label{fig:AxionDetection}
\end{figure}

In Fig.~\ref{fig:AxionDetection}, we show a selection of constraints on $g_{a\gamma\gamma}$ from a range of axion haloscope experiments~\cite{DePanfilis,Hagmann,HAYSTAC:2018rwy,ADMX:2018gho,ADMX:2018ogs,ADMX:2019uok,HAYSTAC:2020kwv,CAPP:2020utb,QUAX:2020adt,CAST:2021rlf,Grenet:2021vbb,ADMX:2021abc}. In addition, we also show constraints from the CAST axion helioscope~\cite{CAST:2017uph} and searches for axion-photon conversion in the magnetospheres around Neutron Stars~\cite{Foster:2020pgt,Battye:2021yue}. Current haloscope experiments typically cover the mass range $m_a \in [2, 40]\,\mu\mathrm{eV}$, while a number of proposed experiments (such as ALPHA~\cite{Lawson:2019brd}, MADMAX~\cite{Beurthey:2020yuq}, ORGAN~\cite{McAllister:2017lkb} and future iterations of ADMX) will target the broader mass range of $m_a \in [0.7, 400]\,\mu\mathrm{eV}$. 

Figure~\ref{fig:AxionDetection} shows that the KSVZ axion is within reach of current and planned haloscope experiments over a wide range of masses. The KSVZ axion has already been excluded by ADMX~\cite{ADMX:2021abc} in the range $2 - 4\,\mu\mathrm{eV}$, while the proposed ALPHA, 
MADMAX and ORGAN (as well as future iterations of ADMX) should cover the range $0.7 - 400\,\mu\mathrm{eV}$. The detection of axions in such experiments would not rely on the new couplings introduced by the Dark Axion Portal, but instead would provide a strong hint at the value of $f_a$, through the relationship between $f_a$ and $m_a$. This in turn would help to pin down the possible parameter space for the Dark Axion portal and the associated production of the dark photon component.

\subsection{Dark photon detection}

In the case of eV-scale dark photons, we focus on the prospects for detection in direct detection experiments, which may search for the absorption of dark photons by electrons, resulting in an excess of ionised electrons with energies $E \sim \mDP$~\cite{DAMIC:2019dcn,XENON:2019gfn,SENSEI:2020dpa,XENON:2020rca,FUNKExperiment:2020ofv}.
Even if the dark photons we consider here are relativistic at the time of structure formation, they are sufficiently heavy so as to be non-relativistic today, and will be detectable as any standard cold DM particle. 
The absorption cross section of a dark photon $\sigma_{\gamma\prime}(m_{\gamma\prime})$ with velocity $v$ is directly related with the photoelectric cross section $\sigma_\gamma$ for a photon with energy $m_{\gamma\prime}c^2$ as~\cite{An:2014twa,Hochberg:2016sqx,Bloch:2016sjj}:
\begin{equation}
    \sigma_{\gamma\prime}(m_{\gamma\prime})v=\epsilon^2\sigma_\gamma(m_{\gamma\prime}c^2)c \,.
\end{equation}
We obtain the values of the photoelectric cross section in silicon above 10 eV from Ref.~\cite{Yeh:1985fij} and below 10 eV from Refs.~\cite{doi:10.1063/1.4923379,PhysRevB.27.985}. 
This allows us to determine the dark photon absorption rate in a target as:
\begin{equation}
    \frac{\diff R}{\diff E_e} =\frac{f_{\gamma'}\rho_\mathrm{DM}}{m_{\gamma\prime}}\epsilon^2\sigma_\gamma(m_{\gamma\prime}c^2)c \,\delta(E_e - m_{\gamma'}) \,,
    \label{eq:rateabsobedDP}
\end{equation}
where $\rho_\mathrm{DM}$ is the local density of DM (for which we assume $\rho_{DM}=0.45 \,\mathrm{GeV}\, \mathrm{cm}^{-3}$~\cite{Read:2014qva}) and $f_{\gamma'}$ is the fraction of DM in dark photons.

We focus here on DAMIC-M (Dark Matter in CCDs at Modane)~\cite{damicm:castello}, a near-future experiment aiming to search for low-mass DM particles through their interactions with silicon atoms in the bulk of charge-coupled devices (CCDs). The use of a CCD for the DM search is based on the creation of electron-hole pairs due to nuclear or electronic recoils in the silicon bulk. 
The CCDs allow for a 3D reconstruction of the energy deposits and the identification of particle types based on the cluster pattern. This technique has been demonstrated by DAMIC, the predecessor experiment located at SNOLAB, with a target mass of around $40\,\mathrm{g}$~\cite{DAMIC:2019dcn,damicm:wimp}.

DAMIC-M will be located at the Modane Underground Laboratory (France) and will have a detector mass around 25 times larger than DAMIC ($1\,\mathrm{kg}$), consisting of 200 CCDs.
These CCDs will be equipped with `skipper' amplifiers which will perform repeated, non-destructive measurements of the charge in each pixel, in order to damp the readout noise 
to a sub-electron level of the order of 0.07 electrons, as has already been demonstrated by the SENSEI experiment~\cite{SENSEI:2020dpa}.  DAMIC-M has a target radiogenic background rate of $\sim 0.1 \, \mathrm{events/keV/kg/day}$, 100 times lower than that of DAMIC. We will also consider the sensitivity of the Low Background Chamber (LBC), a test and development chamber which will be used to characterise DAMIC-M CCDs. The LBC is expect to start taking data in early 2022, with an expected radiogenic background rate of $\sim 3 \, \mathrm{events/keV/kg/day}$~\cite{JuanCortabitarte_Thesis}.

\begin{figure}[t]
    \centering
\includegraphics[width=0.47\textwidth]{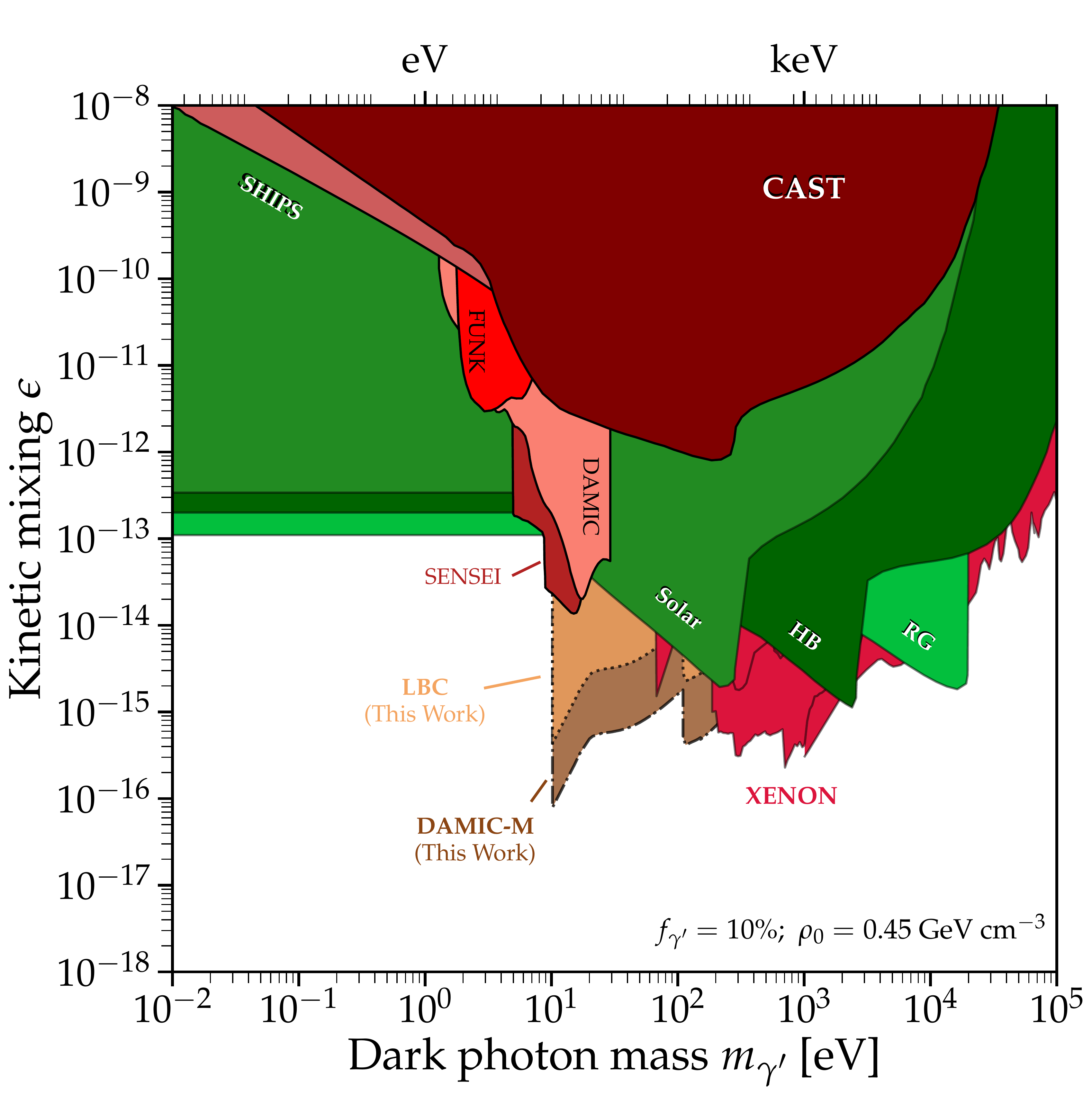}
    \caption{\textbf{Projected constraints on dark photons}, assuming $f_{\DP} = 10\%$. Green regions show constraints from stellar cooling in the Sun, Horizontal Branch (HB) stars and Red Giants (RG)~\cite{Ahlers:2008qc,An:2013yua,An:2020bxd}. Constraints from several direct detection experiments are shown in red: FUNK~\cite{FUNKExperiment:2020ofv}, DAMIC~\cite{DAMIC:2019dcn}, SENSEI~\cite{SENSEI:2020dpa}, and XENON~\cite{XENON:2019gfn,XENON:2020rca}. Shown in dotted and dot-dashed brown are projected constraints which we derive here for the LBC and DAMIC-M, which should probe new parameter space in the mass range $\mDP \sim 10 - 200\,\mathrm{eV}$. Constraints adapted from Ref.~\cite{Caputo:2021eaa}.}
    \label{fig:DarkPhotonDetection}
\end{figure}

In order to estimate the sensitivity of future detectors to dark photon DM, we compare the rate of dark photon absorption events $\diff R/\diff E_e$ with the predicted background. When the absorption rate exceeds the background, we assume that the dark photon will be detectable in future detectors. When the number of electron-hole pairs produced per absorption becomes small, the constraint will depend sensitively on the level of leakage current in the detector~\cite{DAMIC:2016qck}. We therefore conservatively cut off our projected sensitivities below $\mDP < 10 \,\mathrm{eV}$, which would correspond to the creation of roughly 3 electron-hole pairs in the CCD.

Our estimate of the projected bounds obtained in this way are shown in Fig.~\ref{fig:DarkPhotonDetection}, assuming that the dark photon makes up 10\% of the total DM density. DAMIC-M (brown dot-dashed) should constrain new parameter space in the mass range $\mDP \sim 10 - 200\,\mathrm{eV}$, for kinetic mixings as small as $\epsilon \sim 10^{-16}$, with LBC (brown dotted) being roughly a factor of 5 less sensitive.

\section{Discussion}
\label{sec:Discussion}

The Dark Axion Portal Model provides a scenario in which the cosmic density of Dark Matter may be comprised of a mixture of axions and dark photons. We have broadly focused on KSVZ axions in the mass range $m_a \sim [10^{-6}, 10^{-3}]\,\mathrm{eV}$ and kinetically mixed dark photons in the mass range $\mDP \sim [1, 1000]\,\mathrm{eV}$.

\begin{figure}[t]
    \centering
    \includegraphics[width=0.48\textwidth]{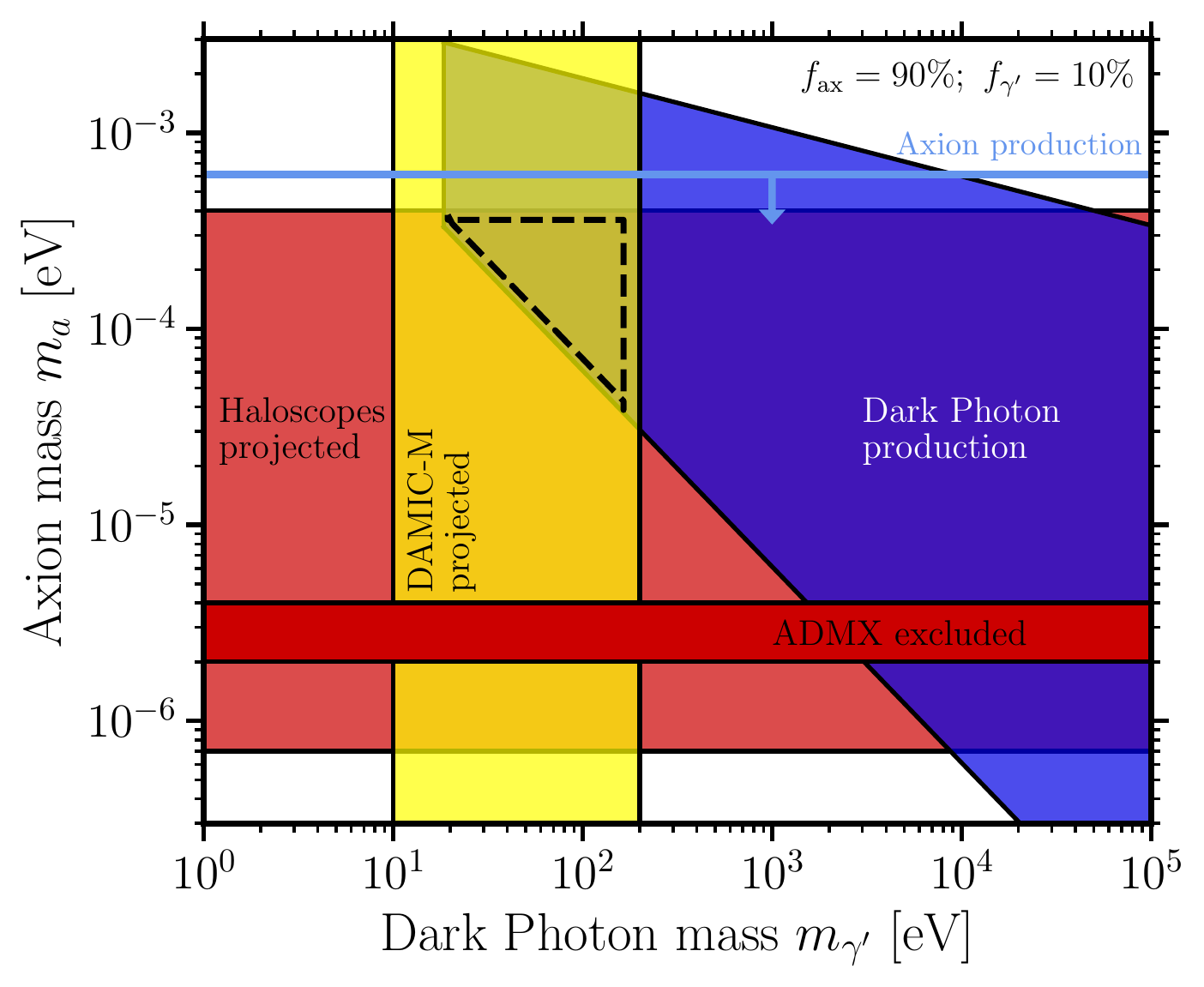}
    \caption{\textbf{Dark Axion Portal parameter space for production and detection of Axions and Dark Photons, assuming $f_\mathrm{ax} = 90\%$ and $f_{\DP} = 10\%$.} We show the allowed regions where dark photons can be produced with $f_{\DP} = 10\%$ and axions can be produced with  $f_\mathrm{ax} = 90\%$ in the pre-inflationary scenario (see Fig.~\ref{fig:FreezeInRegion}). We also show the regions which may be probed with proposed axion haloscopes, some of which is already excluded by ADMX (see Fig.~\ref{fig:AxionDetection}) and the mass range for which DAMIC-M will be able to explore new parameter space for the dark photon kinetic mixing (see Fig.~\ref{fig:DarkPhotonDetection}). The black dashed line highlights the region where both the axion and dark photon can be produced with the correct abundances and may be detectable in planned direct searches.}
    \label{fig:Complementarity_10pct}
\end{figure}

\begin{figure}[t]
    \centering
    \includegraphics[width=0.48\textwidth]{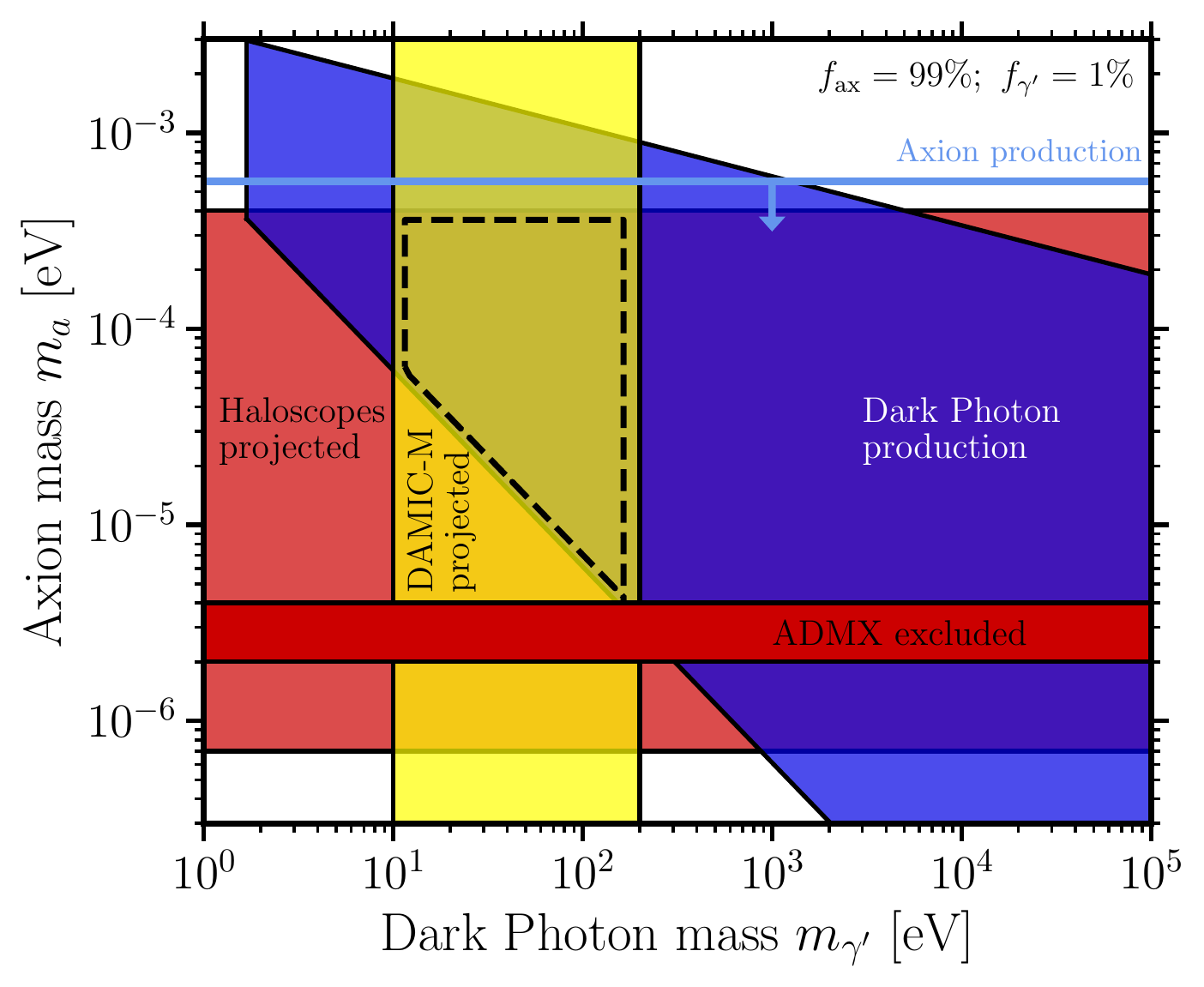}
    \caption{\textbf{Dark Axion Portal parameter space for production and detection of Axions and Dark Photons, assuming $f_\mathrm{ax} = 99\%$ and $f_{\DP} = 1\%$.} The labelled regions are as in Fig.~\ref{fig:Complementarity_10pct}.}
    \label{fig:Complementarity_1pct}
\end{figure}

We have so far explored the possible channels for producing axions and dark photons in the Dark Axion Portal, as well as the prospects for detecting each of these DM candidates. In particular,  in Sec.~\ref{sec:Production}, we have demonstrated that new production channels which arise when we include a non-zero kinetic mixing $\epsilon$ between the visible and dark photons have a negligible impact on the dark photon abundance. Instead, the dark photon density today is set by gluon fusion $gg \rightarrow a \rightarrow \DP\DP$ at high temperatures. Indeed, without a coupling to axions, it seems difficult to produce dark photons through the kinetic mixing alone. However, this `UV freeze-in' mechanism is only possible for sufficiently large re-heating temperatures, $T_\mathrm{RH} \gtrsim 10^9\,\mathrm{GeV}$. If $T_\mathrm{RH}$ is too large, the dark photons enter equilibrium with the Standard Model plasma and the freeze-in mechanism cannot be realised. For a given dark photon abundance, this constraint on $T_\mathrm{RH}$ can be translated into a lower limit on the dark photon mass, and we find that to reach masses in the range $\mDP \sim 1 - 100 \,\mathrm{eV}$ requires that the fraction of DM in dark photons should be less than around $f_{\DP} \lesssim 1\% - 10\%$. Cold axions produced through the misalignment mechanism can make up the remaining $f_\mathrm{ax} \gtrsim 90\% - 99\%$.  
While the light dark photons would potentially act as hot DM, such a `mixed DM' scenario is not yet excluded by cosmological data. 

In Sec.~\ref{sec:Detection}, we explored the ranges of axion and dark photon masses which may be detectable with future experiments. Though the detection of these particles in direct searches does not rely on new couplings introduced in the Dark Axion Portal, the interplay between production and detection of the two candidates allows us to map out the parameter space of the model. We illustrate this in Fig.~\ref{fig:Complementarity_10pct}, for the specific choice of $f_\mathrm{ax} = 90\%$ and $f_{\DP} = 10\%$. We show the ranges of dark photon masses which may be probed with the planned DAMIC-M experiment (yellow) and the ranges of axion masses which will be covered by future axion haloscope searches (red), from ADMX at low masses, to 
ALPHA at high masses. The range of possible masses for producing the desired axion abundance in the pre-inflationary scenario lies below the light blue line (with the precise abundance set by the unknown initial misalignment angle $\theta_i$). The region where dark photons may be produced through freeze-in with the desired abundance is shown in dark blue. In this case, we allow for a range of reheating temperatures $T_\mathrm{RH} > 10^9\,\mathrm{GeV}$, subject to the constraint that $T_\mathrm{RH} < f_a$, in order to preserve the pre-inflationary production of axions as described in Sec.~\ref{sec:Production:Axions}. The cut-off in the `Dark Photon production' region at low $\mDP$ comes from the requirement that the Dark Photons should not enter equilibrium with the Standard Model, as mentioned above.

With these constraints, we can pinpoint the region of parameter space whether both axions and dark photons can be produced in the early Universe \textit{and} detected in future experiments. This region is highlighted with a dashed black line in Fig.~\ref{fig:Complementarity_10pct}, spanning masses in the range $\mDP \sim 20 - 200 \,\mathrm{eV}$ and $m_a \sim 30 - 400 \,\mu\mathrm{eV}$. If instead axions are produced in the post-inflationary scenario, this selects the somewhat narrower range $m_a \sim 20 - 50 \,\mu\mathrm{eV}$ for the axion mass (not shown in Fig.~\ref{fig:Complementarity_10pct}), and a similar range of dark photon masses as in the pre-inflationary case. These parameter ranges therefore provide concrete theoretical targets for direct searches for axion and dark photon DM.

In Fig.~\ref{fig:Complementarity_1pct}, we show similar regions for the ratio $f_\mathrm{ax} = 99\%$ and $f_{\DP} = 1\%$. As shown in Fig.~\ref{fig:FreezeInRegion}, this lower dark photon abundance allows for a wider range of masses to be produced. Even with this smaller abundance set by the Dark Axion Portal, we find that future experiments such as DAMIC-M will still be able explore new regions in the $(\mDP, \epsilon)$ parameter space.

We have focused throughout the paper on a benchmark choice for the couplings of the Dark $U(1)_D$, specifically $e^\prime = 0.1$ and $D_\psi = 3$, assuming that the new heavy fermions have a similar interaction strength under $U(1)_D$ as electrons do under $U(1)_\mathrm{QED}$. As we show in Appendix~\ref{app:Charges}, moving away from this assumption changes the specific mass ranges suggested in Figs.~\ref{fig:Complementarity_10pct}~and~\ref{fig:Complementarity_1pct}, though the complementarity between the different search strategies remains.

The Dark Axion Portal has been gaining attention as a model for mixed Dark Matter. Here, we have performed the first in-depth study of the prospects for direct detection of DM in the Dark Axion Portal. For the dark photon, detection relies crucially on a kinetic mixing $\epsilon$ with the visible photon, which has not previously been studied in detail in this context. We have demonstrated, however, that such a mixing does not substantially alter the stability or relic abundances of either the axion or dark photon over the masses we consider.  We have also illustrated that (depending on the dark photon kinetic mixing) both DM particles may be directly detected in upcoming detectors. Light dark photons may be detected through dark photon absorption, to which CCD-based detectors such as DAMIC-M are sensitive. Axions in the mass range of interest may be detected using axion haloscopes 
. 
If both DM particles are detected, it will indirectly provide us with information about early Universe cosmology which may be otherwise inaccessible, including the energy scale of inflation and the reheating temperature, through the production mechanisms we have discussed here. Crucially, it will also mark a major advance in pinning down models of particle physics beyond the Standard Model. 

\acknowledgements

The authors thank Belen Barreiro, Alicia Calder\'on, Marco Gorghetto, Kunio Kaneta, Marius Millea, Patricio Vielva, Luca Visinelli and Sam Witte for helpful discussions. The authors also thank Sven Heinemeyer and Juan Manuel Socu\'ellamos for useful comments on this manuscript.
The authors are also grateful to Ciaran O'Hare for his maintenance of the public webpage \href{https://cajohare.github.io/AxionLimits/}{cajohare.github.io/AxionLimits/}~\cite{axionlimits} from which a number of digitised constraints were obtained.

The authors also thank the DAMIC-M collaboration for their cooperation.

B.J.K. and F.J.C.\ thank the Spanish Agencia Estatal de Investigaci\'on (AEI, MICIU) for the support to the Unidad de Excelencia Mar\'ia de Maeztu Instituto de F\'isica de Cantabria, ref. MDM-2017-0765. E.M.G. and F.J.C.\ thank the Spanish Agencia Estatal de Investigaci\'on (AEI, MCI) for the funds received through the research project, ref. PID2019-110610RB-C21. J.M.D.  acknowledges  the  support  of  project  PGC2018-101814-B-100 (MCIU/AEI/MINECO/FEDER, UE) Ministerio de Ciencia, Investigación y Universidades.

This project was initiated and supported through the activities of the `Dark Collaboration at IFCA' working group.

\bibliography{references}

\clearpage

\appendix

\section{Impact of Dark Coupling Constant}
\label{app:Charges}

Here, we briefly discuss the impact of changing the Dark $U(1)_D$ coupling constant $e^\prime$ on our results. The fraction of DM in dark photons is very sensitive to the couplings in the Dark Sector, with
\begin{equation}
    f_{\DP} \propto (e^\prime D_\psi)^4 T_\mathrm{RH}^3/f_a{}^4\,,
\end{equation} 
for production through gluon fusion, as shown in Eq.~\eqref{eq:DP_fraction}. In the main text, we focus on the benchmark scenario $e^\prime = 0.1$, $D_\psi = 3$. However, varying these values (at fixed $f_{\DP}$) leads to a shift in the required axion decay constant and reheating temperature.

\begin{figure}[!htb]
\centering
     \includegraphics[width=0.47\textwidth]{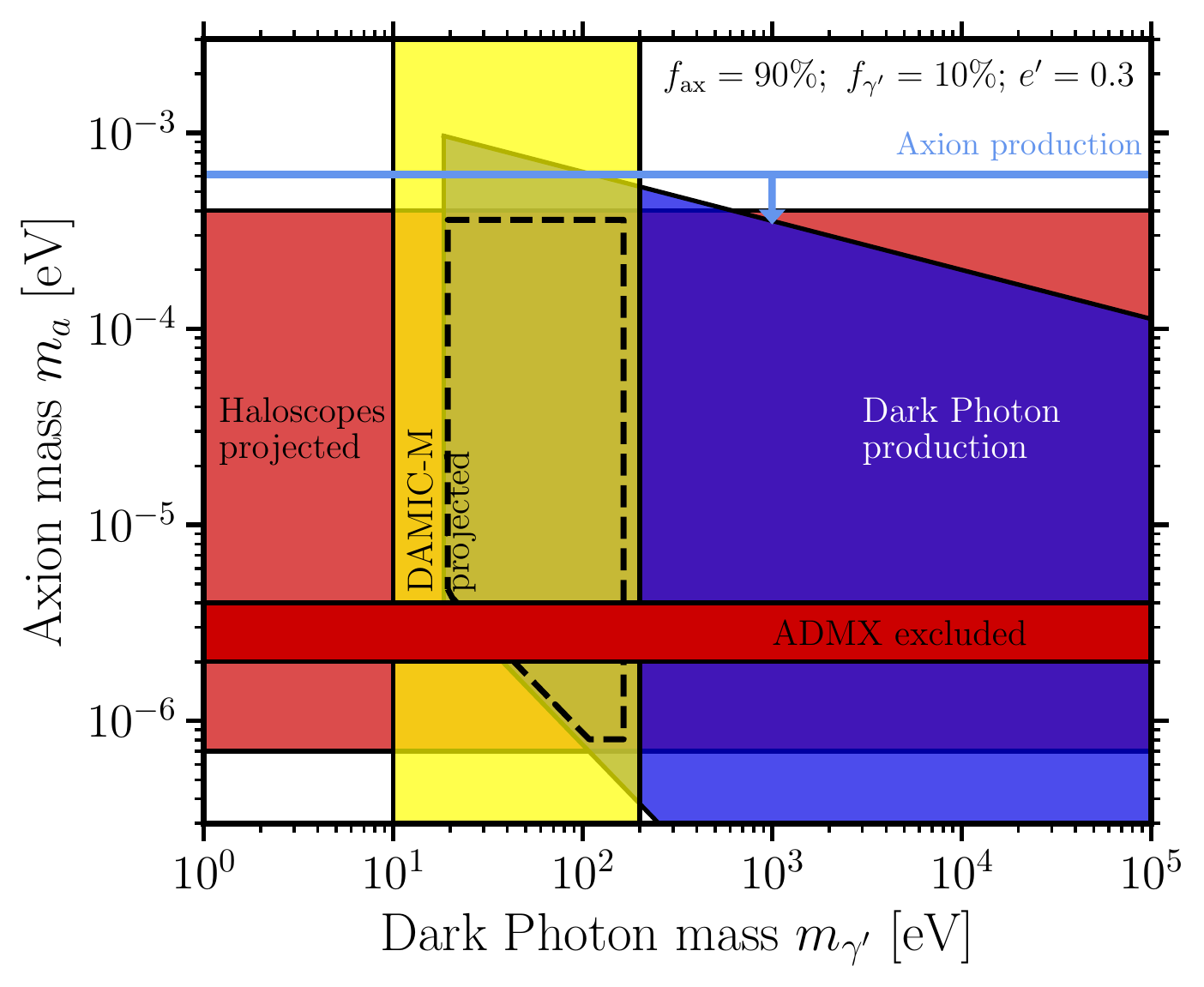}
    \includegraphics[width=0.47\textwidth]{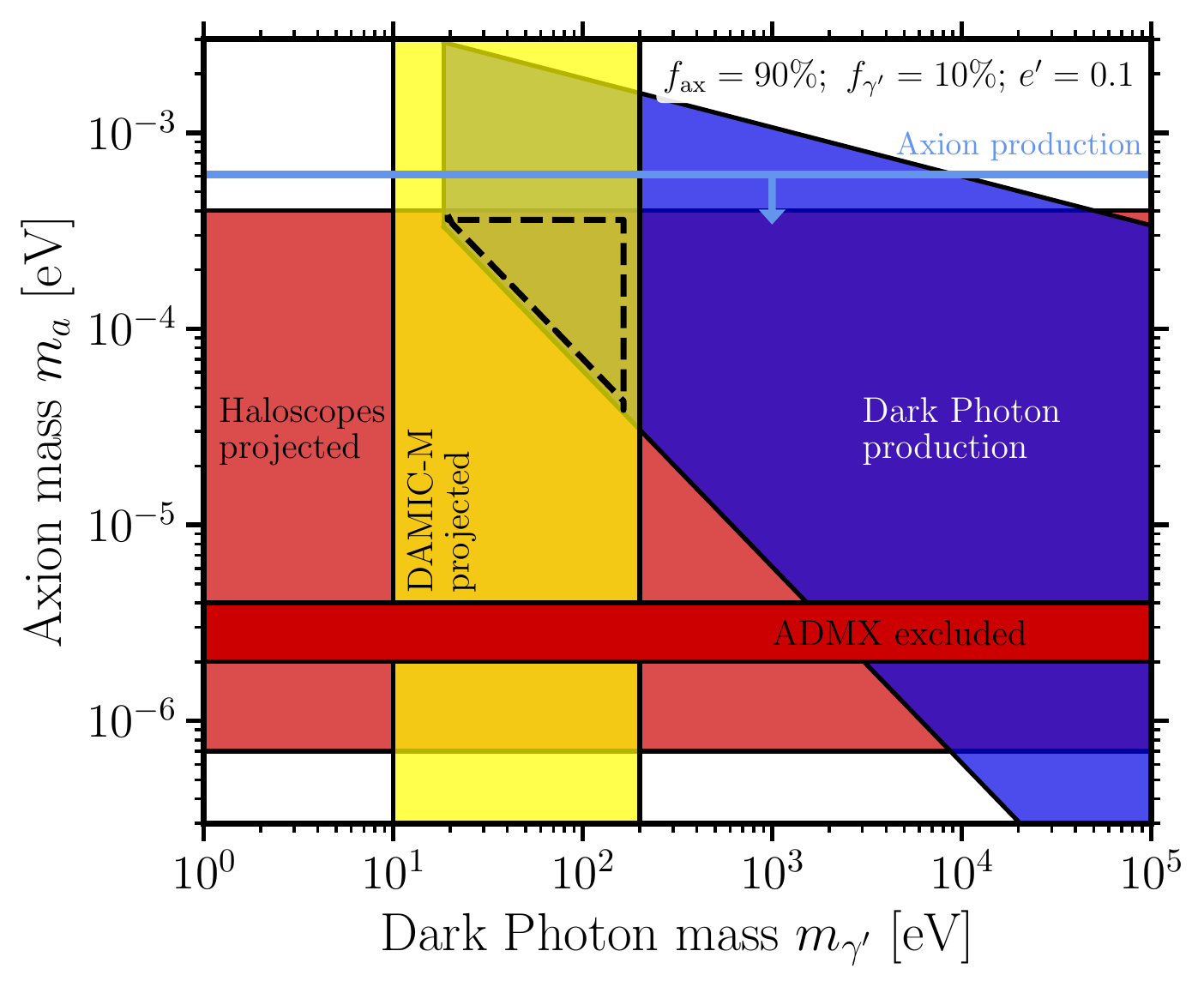}    
    \includegraphics[width=0.47\textwidth]{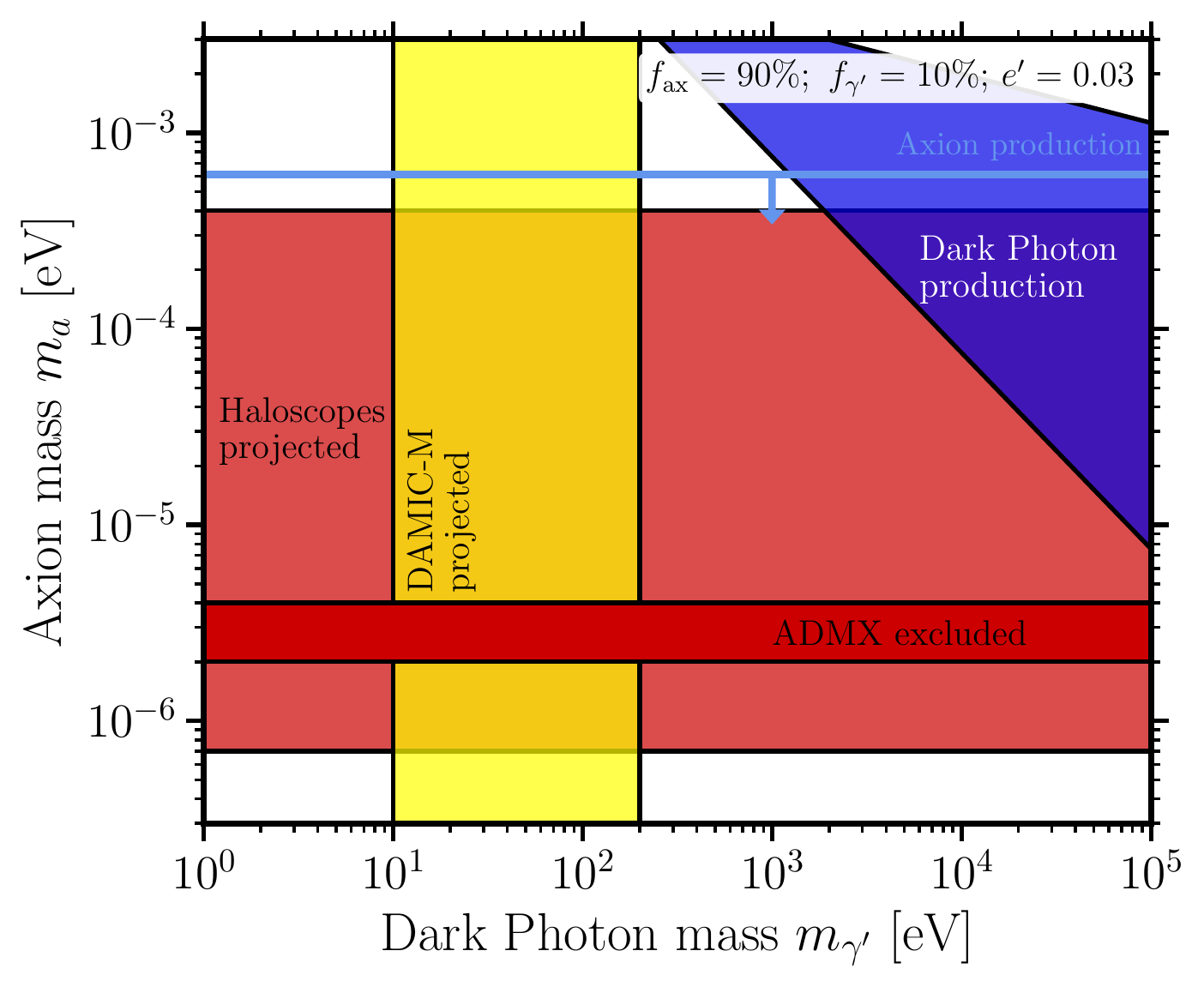}   
    \caption{\textbf{Dark Axion Portal parameter space for production and detection of Axions and Dark Photons, for different values of the Dark $U(1)_D$ Coupling}: $e^\prime = 0.3$ (top), $e^\prime = 0.1$ (middle), $e^\prime = 0.03$ (bottom). In all cases, we assume $f_\mathrm{ax} = 90\%$ and $f_{\DP} = 10\%$. Labelled regions as in Fig.~\ref{fig:Complementarity_10pct} (which appears here as the middle panel).}
    \label{fig:Complementarity_10pct_eD}
\end{figure}

We summarise the Dark Axion Portal parameter space in Fig.~\ref{fig:Complementarity_10pct_eD} for different values of $e^\prime$, assuming $f_{\DP} = 10\%$ as in Fig.~\ref{fig:Complementarity_10pct}. Increasing $e^\prime$ by a factor of 3 (top panel) compared to our fiducial value leads to a shift towards higher $f_a$ and correspondingly lower $m_a$. In this case, the Dark Axion Portal leads to viable production of dark photons and axions over almost the entire range of masses to which proposed haloscopes should be sensitive.

Conversely, reducing the dark coupling to $e^\prime = 0.03$ causes the allowed region for dark photon production to shrink. Smaller values of $e^\prime$ require large values of $T_\mathrm{RH}$ to achieve the same dark photon density. This in turn means that the condition $T_\mathrm{RH} > f_a$ (required for pre-inflationary axion production) is not satisfied over a wide range of the parameter space. This constraint sets the lower boundary of the `Dark Photon production' region, below which axions can only be produced in the post-inflationary scenario. Of course, in this scenario, the Dark Axion Portal can still be used to produce detectable dark photons and axions, though only in the narrower axion mass range $m_a \sim 20 - 50 \,\mu\mathrm{eV}$, as suggested by simulations~\cite{Buschmann:2019icd}.

These examples illustrate that the arguments in the main text for the production of axions in the pre-inflationary scenario are only valid for dark sector coupling sufficiently close to the Standard Model electromagnetic coupling (within a factor of a few). \textit{A priori}, these couplings are unknown, but even if they are much smaller than in the Standard Model, axion production may still be possible if the Peccei-Quinn symmetry is broken after inflation.

\FloatBarrier

\end{document}